\documentclass[11pt]{article}

\usepackage[final]{acl}


\usepackage{array}
\usepackage{times}
\usepackage{latexsym}
\usepackage{booktabs}
\usepackage{float}
\usepackage{amsmath}
\usepackage{algorithm}
\usepackage{multirow}
\usepackage{algpseudocode}
\usepackage{adjustbox}
\usepackage{makecell}

\usepackage{listings}
\usepackage{xcolor}
\usepackage{enumitem}
\usepackage{needspace}

\newenvironment{prompttext}{\par}{\par}

\newenvironment{promptbox}[1]{%
  \par\noindent\textbf{#1}\par\nobreak
}{%
  \par
}

\setlist[itemize]{leftmargin=1.6em}
\setlist[enumerate]{leftmargin=1.8em}

\lstdefinestyle{jsonstyle}{
    basicstyle=\footnotesize\ttfamily,
    breaklines=true,
    columns=fullflexible,
    keepspaces=true,
    frame=none,
    backgroundcolor=\color{gray!10},
    showstringspaces=false
}

\usepackage[T1]{fontenc}

\usepackage[utf8]{inputenc}

\usepackage{microtype}

\usepackage{inconsolata}

\usepackage{graphicx}

\title{CPM-MultiAgent: A CPM-Grounded Appraisal Framework for Simulating Dynamic Patient Emotional Evolution in Healthcare Dialogue}

\author{
\textbf{Jingyao Cai\textsuperscript{1}, Shuaijun Liu\textsuperscript{2}, Xiaowan Li\textsuperscript{3}, Yutong Guo\textsuperscript{1},} \\
\textbf{Abdul Rehman\textsuperscript{1}, Qintian\textsuperscript{4}, Thomas Dolby\textsuperscript{5}, Sue M Green\textsuperscript{1}, Xiaosong Yang\textsuperscript{1}} \\
{\normalfont \textsuperscript{1}Bournemouth University, Bournemouth, UK} \\
{\normalfont \textsuperscript{2}HKUST(GZ), Guangzhou, China\quad \textsuperscript{3}Hiroshima University, Hiroshima, Japan} \\
{\normalfont \textsuperscript{4}Qiongtai Normal University, Haikou, China\quad \textsuperscript{5}i3 Simulations Ltd, UK} \\
{\normalfont\texttt{\{jcai, s5819307, arehman, sgreen, xyang\}@bournemouth.ac.uk}} \\
{\normalfont\texttt{sliu529@connect.hkust-gz.edu.cn, d233196@hiroshima-u.ac.jp}} \\
{\normalfont\texttt{qintian@mail.qtnu.edu.cn, tom@i3simulations.com}}
}

\begin{document}

\maketitle
\begin{abstract}
Large Language Models (LLMs) enable patient simulation for healthcare communication training, where learners practice responding to clinical information, concerns, and emotional reactions. However, existing patient simulation and affective dialogue systems often represent patient emotions as static attributes or focus on recognizing emotions and generating empathetic responses. They provide limited support for modeling how a simulated patient's emotional state evolves in response to a healthcare professional's communication. Changes in patient emotion constitute interactional feedback that learners must recognize and respond to across dialogue turns. We draw on the Component Process Model (CPM), a psychological theory that explains emotion as a dynamic process shaped by event appraisal. We propose CPM-MultiAgent, a CPM-grounded multi-agent framework for dynamic patient emotion simulation in healthcare communication training. The framework uses an emotion taxonomy informed by medical literature and expert input, and updates emotional states through healthcare-related trigger identification, structured CPM appraisal, and emotion-state integration, accounting for the patient's profile, clinical context, and prior emotional state. We evaluate the framework through student-led conversations and controlled trigger cases, combining nursing-expert assessments, LLM-judge comparisons, component ablations, backbone checks, and student feedback.

\end{abstract}

\section{Introduction}

LLMs are increasingly used to simulate patients for healthcare communication training~\citep{wang2024patient}. In nursing education, these systems provide opportunities to practice responding to patient concerns and managing challenging conversations. Realistic patient simulation requires the virtual patient's responses to reflect the nursing student's communication, including changes in the patient's cooperation, resistance, and emotional expression~\citep{lee2025adaptive}. Modeling these changes is therefore important for providing meaningful interactional feedback during communication practice.

\begin{figure*}[!t]
  \centering
  \includegraphics[width=0.95\textwidth]{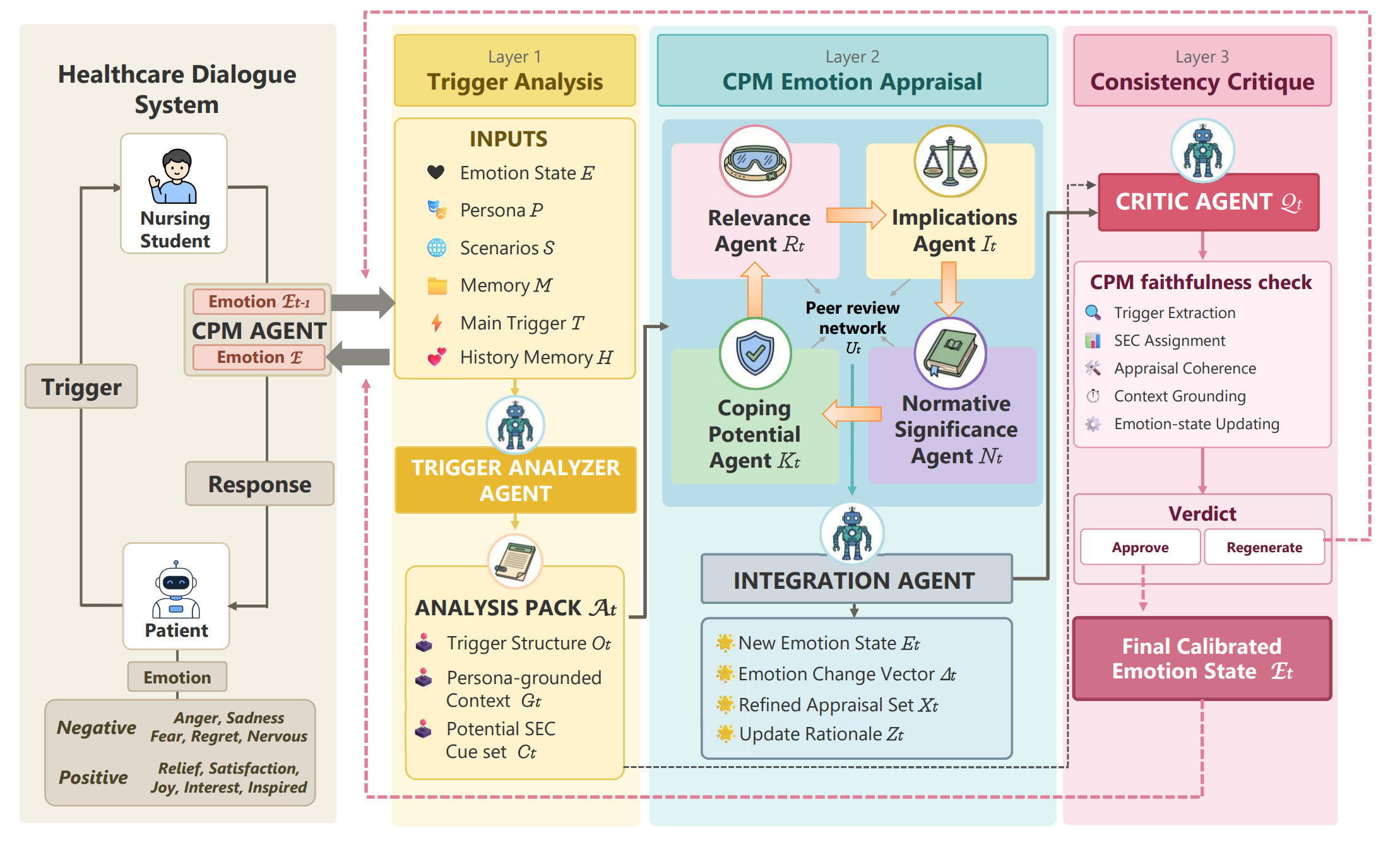}
  \caption{Overview of the proposed CPM-MultiAgent framework for dynamic patient emotion simulation in healthcare communication training. Stimulus evaluation check (SEC) cues inform the appraisal agents.}
  \label{fig:framework1}
\end{figure*}

A central challenge is to make patient reactions both responsive to communication and consistent with the patient's individual context. Evaluations have found that LLM-based virtual patients can be overly agreeable and may not react appropriately when clinicians communicate poorly~\citep{cook2025virtual}. Recent systems have made patient behavior and internal states responsive to the interaction~\citep{steenstra2025scaffolding,fajardo2026medpi}. However, adaptive responses and updated emotion scores do not, by themselves, explain why a particular utterance changes a patient's emotional state. Appraisal theory links emotional responses to how individuals evaluate events against their concerns, goals, and coping resources. For nursing communication training, the modeling challenge is to make this connection explicit while preserving the patient's clinical context and emotional history. We therefore ask how dialogue triggers can be translated into patient-specific appraisals that support plausible and temporally coherent emotional transitions.

To address this question, we draw on CPM~\citep{scherer2009dynamic}, which describes emotion as a dynamic process arising from structured event appraisal. Its four appraisal checks cover relevance, implications, coping potential, and normative significance. We propose CPM-MultiAgent, a multi-agent framework that uses this structure to model patient emotional transitions in healthcare communication training (Figure~\ref{fig:framework1}). A Trigger Analyzer Agent identifies salient dialogue events, and four appraisal agents evaluate their significance for the patient. An Integration Agent updates the emotional state, while a Critic Agent checks the transition for contextual consistency, temporal coherence, and CPM faithfulness. This division makes the reasoning between a communicative event and an emotional update explicit.

The study makes three contributions to patient simulation for healthcare communication training. First, we define emotion categories informed by medical literature and expert input for patient simulation in healthcare communication training. Second, CPM-MultiAgent connects dialogue triggers to emotional updates through patient-specific appraisal, state integration, and consistency critique. Third, we evaluate transition plausibility, contextual grounding, temporal consistency, and appraisal quality through baseline comparisons, ablations, model-based assessment, and case analyses. The evaluation also includes assessments by nursing experts and students.

\section{Related Works}

\subsection{Virtual Patient Simulation}
LLM-based virtual patient systems differ in how they represent patient characteristics, coordinate simulation components, and provide feedback to trainees. \citet{cook2025virtual} examined dialogue authenticity, the representation of patient preferences, and personalized feedback in LLM-based patient simulation. Psychologically grounded approaches include PATIENT-$\Psi$~\citep{wang2024patient}, which conditions simulated patients on cognitive behavioral therapy models, and SimPatient~\citep{steenstra2025scaffolding}, which uses multiple LLM agents for motivational interviewing practice, utterance-level feedback, and dynamic updates to patient cognitive factors. For clinical communication training, \citet{qu2026multiagent} coordinate specialized agents for character mapping, memory management, information processing, language generation, and response checking. Most directly related to nursing education, Adaptive-VP~\citep{lee2025adaptive} combines multi-agent assessment of trainee communication with score-guided changes in patient cooperation, complaint intensity, and conversational behavior. Together, these approaches emphasize cognitive fidelity, reliable role enactment, and adaptation to trainee performance. CPM-MultiAgent complements these goals by explicitly modeling the appraisal mechanism governing patient emotion updates.

\subsection{Dynamic Emotion Modeling}
Recent LLM-based systems explicitly maintain or regulate affect across dialogue turns. In psychological counseling, AnnaAgent~\citep{wang2025annaagent} combines an emotion modulator, a complaint elicitor, and multi-session memory to simulate evolving seeker states. In medical dialogue evaluation, MedPI~\citep{fajardo2026medpi} uses a dedicated emotion-state LLM to update a patient's 27-dimensional affective representation after each clinician turn and couples these updates with memory retrieval. Beyond healthcare, Sentipolis~\citep{fu2026sentipolis} integrates continuous pleasure--arousal--dominance states, dual-speed emotion dynamics, and emotion--memory coupling for social simulation. Policy-oriented approaches, including EQ-Negotiator~\citep{liu2025eq} and EvoEmo~\citep{long2025evoemo}, regulate emotional expression through game-theoretic and hidden Markov policies or evolutionary optimization of a Markov decision process for negotiation. Complementing these state- and policy-based approaches, the Third-Person Appraisal Agent~\citep{hong2025third} applies primary appraisal, secondary appraisal, and reappraisal to emotional reasoning, with generator--evaluator feedback and reinforced fine-tuning. These approaches target seeker simulation, medical dialogue evaluation, social continuity, negotiation performance, or emotional inference. Our framework applies structured CPM appraisal to the simulated patient's own emotional transitions during nursing dialogue.

\section{Methodology}

CPM-MultiAgent updates patient emotion during nursing communication through trigger analysis, CPM appraisal and integration, and consistency critique (Figure~\ref{fig:framework1}). A patient-oriented representation defines the updated states (Section~\ref{sec:patient_emotion_representation}), while CPM links dialogue events to emotional changes (Section~\ref{sec:cpm}). The patient's profile, clinical context, dialogue history, and prior emotion state inform each update. The Critic Agent checks CPM faithfulness, contextual grounding, and temporal coherence before the updated state guides response generation.

\subsection{Patient Emotion Representation}
\label{sec:patient_emotion_representation}
We adapt the eight emotion categories used by \citet{pilet2025feelings} to study patients admitted to emergency departments. Their adapted Geneva Emotion Wheel comprises four negative emotions (\textit{anger}, \textit{fear}, \textit{regret}, and \textit{sadness}) and four positive emotions (\textit{relief}, \textit{satisfaction}, \textit{joy}, and \textit{interest}). To extend this representation to nursing communication, we discussed the categories with nursing experts and added \textit{nervous} to the negative group and \textit{inspired} to the positive group. These additions capture states the experts considered relevant to routine nurse--patient interactions: tension associated with uncertainty or interpersonal pressure, and strengthened hope, agency, confidence, or motivation. The resulting representation contains ten emotion categories, with five in each valence group.

Through these discussions, we also specified nursing-specific criteria for interpreting each emotion in relation to the patient's situation and the current communicative event. For example, \textit{fear} concerns an identifiable threat perceived as credible, whereas \textit{nervous} concerns uncertainty, waiting, unpredictability, or unconfirmed adverse possibilities. Appendix~\ref{sec:patient_emotion_criteria} provides the criteria for all ten categories. These criteria operationalize our adapted representation for nursing communication; they are distinct from the eight emotion categories adopted from that study. Each emotion receives a separate 1--5 intensity score, allowing multiple positive and negative emotions to coexist. The intensity anchors are given in Table~\ref{tab:emotion_intensity}.

\subsection{CPM-Grounded Emotion Modeling}
\label{sec:cpm}
\begin{figure}[!t]
  \centering
  \includegraphics[width=\columnwidth]{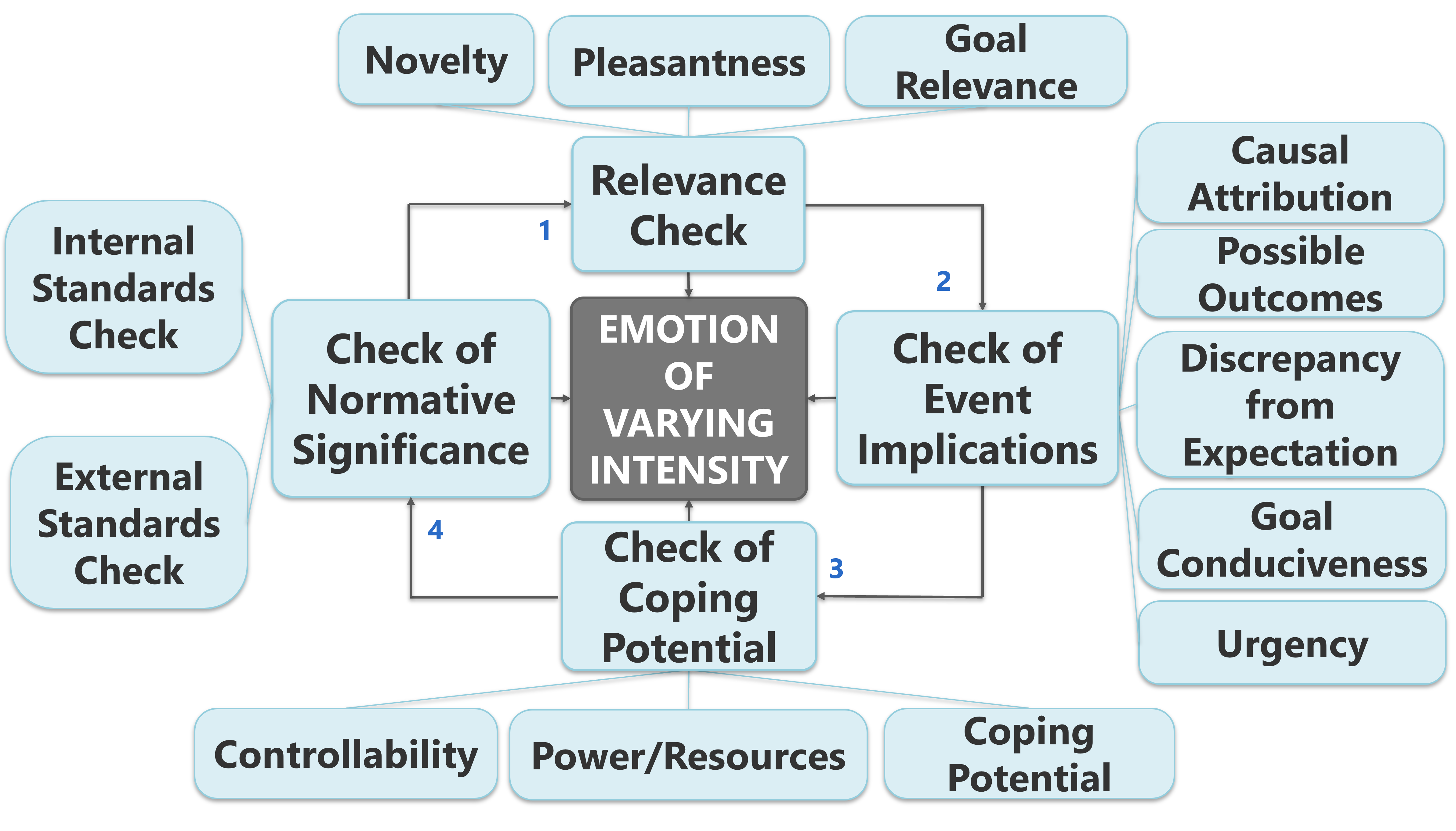}
  \caption{CPM appraisal structure used in our framework.}
  \label{fig:cpm}
\end{figure}

CPM views emotion as a dynamic process arising from the appraisal of events or stimuli, rather than as a fixed affective label. In CPM, the same event may lead to different emotions depending on how it is evaluated with respect to an individual's concerns, goals, coping conditions, and norms. The appraisal process is organized into four major checks. \textbf{Relevance} determines whether a stimulus is novel, pleasant or unpleasant, and relevant to the patient's goals. \textbf{Implication} estimates the stimulus's causes, expectedness, likely outcomes, goal conduciveness, and urgency. \textbf{Coping potential} assesses whether the patient can control, adapt to, or cope with the consequences. \textbf{Normative significance} evaluates whether the event is compatible with the patient's self-concept and social norms. The checks and finer-grained SECs are adapted from \citet{scherer2009dynamic} and \citet{yarwood2022psychology} (Table~\ref{tab:cpm}; Appendix~\ref{sec:cpm_appendix}). Figure~\ref{fig:cpm} summarizes this appraisal structure.

In nursing communication, a treatment explanation may elicit different emotions depending on the patient's concerns, expectations, and perceived control. The patient profile, clinical context, dialogue history, and prior emotional state therefore inform appraisal. Our agents apply complementary CPM checks to this context and integrate their results into updates that allow multiple patient emotions to coexist at different intensities.

\subsection{Trigger Analysis Layer}

The Trigger Analysis Layer structures events in the nurse's utterance and recent dialogue, such as a procedure explanation or a response to patient concerns. It grounds these events in the patient's profile and clinical context and extracts CPM-related appraisal cues without directly inferring the new emotional state.

\textbf{Trigger Analyzer Agent.} At dialogue turn \(t\), the Trigger Analyzer Agent takes as input the patient profile \(P\), clinical context \(S\), dialogue memory \(M\), recent dialogue history \(H_t\), previous emotion state \(E_{t-1}\), and main trigger \(T_t\). It outputs an analysis pack \(A_t\):
\begin{equation}
  \label{eq:trigger-analyzer}
  A_t =
  \mathrm{TriggerAnalyzer}(P, S, M, H_t, E_{t-1}, T_t).
\end{equation}

The analysis pack contains three parts. The objective trigger structure \(O_t\) records the communicative event, its participants, and its position in the dialogue. The patient-specific context \(G_t\) relates this event to the patient's concerns, goals, relationships, clinical situation, memory, and prior emotional state. The potential SEC cue set \(C_t\) organizes the resulting appraisal cues according to the CPM structure.
\begin{equation}
  \label{eq:analysis-pack}
  A_t = \{O_t, G_t, C_t\}.
\end{equation}

Following the CPM structure illustrated in Figure~\ref{fig:cpm}, \(C_t\) is organized into four appraisal-check groups. Relevance \(C_t^{\mathrm{rel}}\) includes novelty, intrinsic pleasantness, and goal relevance. Implication \(C_t^{\mathrm{imp}}\) includes causal attribution, outcome probability, expectation discrepancy, goal conduciveness, and urgency. Coping potential \(C_t^{\mathrm{cop}}\) includes control, power, and adjustment. Normative significance \(C_t^{\mathrm{norm}}\) includes internal and external standards:
\begin{equation}
  \label{eq:sec-cue-structure}
  \begin{aligned}
  C_t = \{&
  C_t^{\mathrm{rel}}=\{c_{t,\mathrm{rel}}^{1},c_{t,\mathrm{rel}}^{2},c_{t,\mathrm{rel}}^{3}\},\\
  & C_t^{\mathrm{imp}}=\{c_{t,\mathrm{imp}}^{1},\ldots,c_{t,\mathrm{imp}}^{5}\},\\
  & C_t^{\mathrm{cop}}=\{c_{t,\mathrm{cop}}^{1},c_{t,\mathrm{cop}}^{2},c_{t,\mathrm{cop}}^{3}\},\\
  & C_t^{\mathrm{norm}}=\{c_{t,\mathrm{norm}}^{1},c_{t,\mathrm{norm}}^{2}\}
  \}.
  \end{aligned}
\end{equation}

In this way, the Trigger Analyzer Agent transforms raw dialogue input into a CPM-oriented trigger representation, while leaving appraisal and emotion updating to the following multi-agent appraisal layer.

\subsection{CPM Emotion Appraisal Layer}

The CPM Emotion Appraisal Layer evaluates \(A_t=\{O_t,G_t,C_t\}\) from the patient's perspective through four appraisal agents and an Integration Agent. The agents assess relevance, implications, coping potential, and normative significance using the CPM questions and SECs in Appendix~\ref{sec:cpm_appendix}. Later checks condition on preceding results and may revise earlier interpretations through peer review. The Integration Agent combines the refined appraisals to update the patient's emotional state.

\textbf{Relevance Appraisal Agent.}
The Relevance Appraisal Agent evaluates whether the trigger matters to the patient's concerns, care goals, or significant others. It takes \(O_t\), \(G_t\), and \(C_t^{\mathrm{rel}}\) as input, and produces the relevance appraisal result \(R_t\):
\begin{equation}
  \label{eq:rel-agent}
  R_t = \mathrm{RelAgent}(O_t, G_t, C_t^{\mathrm{rel}}).
\end{equation}

\textbf{Implication Appraisal Agent.}
The Implication Appraisal Agent evaluates the trigger's expected consequences for the patient's well-being and short- or long-term care goals. It takes \(O_t\), \(G_t\), \(C_t^{\mathrm{imp}}\), and \(R_t\) as input, and produces the implication appraisal result \(I_t\):
\begin{equation}
  \label{eq:imp-agent}
  I_t = \mathrm{ImpAgent}(O_t, G_t, C_t^{\mathrm{imp}}, R_t).
\end{equation}

\textbf{Coping Potential Appraisal Agent.}
The Coping Potential Appraisal Agent evaluates the patient's perceived control and resources for coping with the anticipated consequences. It takes \(O_t\), \(G_t\), \(C_t^{\mathrm{cop}}\), \(R_t\), and \(I_t\) as input, and produces the coping potential appraisal result \(K_t\):
\begin{equation}
  \label{eq:cop-agent}
  K_t = \mathrm{CopAgent}(O_t, G_t, C_t^{\mathrm{cop}}, R_t, I_t).
\end{equation}

\textbf{Normative Significance Appraisal Agent.}
The Normative Significance Appraisal Agent evaluates compatibility with the patient's self-concept, values, and expectations of respectful care, considering internal standards and social norms. It takes \(O_t\), \(G_t\), \(C_t^{\mathrm{norm}}\), \(R_t\), \(I_t\), and \(K_t\) as input, and produces the normative significance appraisal result \(N_t\):
\begin{equation}
  \label{eq:norm-agent}
  N_t = \mathrm{NormAgent}(O_t, G_t, C_t^{\mathrm{norm}}, R_t, I_t, K_t).
\end{equation}

\textbf{Peer Review.}
After the four agents complete their initial appraisals, they conduct peer review to check the coherence of the appraisal chain. The initial appraisal set is:
\begin{equation}
  \label{eq:initial-appraisal-set}
  X_t = \{R_t, I_t, K_t, N_t\}.
\end{equation}

During peer review, the agents examine whether the appraisal results are consistent with the trigger, patient-specific context, and previous emotional state. The refined appraisal set is:
\begin{equation}
  \label{eq:peer-review}
  \widetilde{X}_t = \mathrm{PeerReview}(X_t, O_t, G_t, E_{t-1}).
\end{equation}
This process allows later checks to challenge or revise earlier ones, reflecting the dynamic and recursive structure of CPM.

\textbf{Integration Agent.}
The Integration Agent updates the patient's emotional state from the refined appraisals. For nursing simulation, \(\mathcal{Y}=\{y_1,y_2,\ldots,y_K\}\) comprises the ten categories in Section~\ref{sec:patient_emotion_representation}, interpreted using Appendix~\ref{sec:patient_emotion_criteria}. The previous state records each label's intensity:
\begin{equation}
  \label{eq:previous-emotion-state}
  E_{t-1}=\{(y_k,e_{t-1,k})\}_{k=1}^{K}.
\end{equation}

The Integration Agent takes \(\widetilde{X}_t\), \(E_{t-1}\), and \(\mathcal{Y}\) as input, and produces an emotion change vector \(\Delta_t\):
\begin{equation}
  \label{eq:emotion-delta}
  \Delta_t = \mathrm{EmotionDelta}(\widetilde{X}_t,E_{t-1},\mathcal{Y}).
\end{equation}

For each label \(y_k\), the updated intensity is clipped to the 1--5 range:
\begin{equation}
  \label{eq:emotion-update}
  e_{t,k} = \mathrm{clip}(e_{t-1,k}+\Delta_{t,k}).
\end{equation}

The final output \(U_t\) contains the updated emotion state \(E_t\), emotion change vector \(\Delta_t\), refined appraisal set \(\widetilde{X}_t\), and update rationale \(Z_t\):
\begin{equation}
  \label{eq:emotion-layer-output}
  U_t = \{E_t, \Delta_t, \widetilde{X}_t, Z_t\}.
\end{equation}

\subsection{Consistency Critique Layer}
\textbf{Critic Agent.}
The Consistency Critique Layer reviews trigger analysis and emotion appraisal results before response generation. At turn \(t\), the original context is denoted as \(\Omega_t\), containing the patient profile \(P\), clinical context \(S\), dialogue memory \(M\), recent dialogue history \(H_t\), previous emotion state \(E_{t-1}\), and main trigger \(T_t\):
\begin{equation}
  \label{eq:context-bundle}
  \Omega_t=\{P,S,M,H_t,E_{t-1},T_t\}.
\end{equation}

The Critic Agent takes \(\Omega_t\), the analysis pack \(A_t\), and the emotion-layer output \(U_t\), and produces a critique decision \(Q_t\):
\begin{equation}
  \label{eq:critic-agent}
  Q_t = \mathrm{CriticAgent}(\Omega_t,A_t,U_t).
\end{equation}
It checks CPM faithfulness, appraisal coherence, and consistency with the patient's clinical context and prior state. For example, reassurance alone need not justify immediate relief if the patient's concern remains unresolved. The decision contains a verdict \(v_t\), target layer \(\ell_t\), and rationale \(b_t\):
\begin{equation}
  \label{eq:critique-decision}
  Q_t=\{v_t,\ell_t,b_t\}.
\end{equation}
Here, \(v_t\) indicates whether the output is approved or requires regeneration, \(\ell_t\) specifies the layer to revise, and \(b_t\) records the rationale.

As summarized in Algorithm~\ref{alg:critic}, if \(v_t=\mathrm{approve}\), the current output is accepted. Otherwise, \(b_t\) is sent back to the target layer for revision and rechecked, with the process bounded by a maximum revision budget \(R\).

\begin{algorithm}[t]
\caption{Consistency Critique Layer}
\label{alg:critic}
\begin{algorithmic}[1]
\Require Original context \(\Omega_t\), analysis pack \(A_t\), emotion-layer output \(U_t\), maximum revision rounds \(R\)
\Ensure Final emotion-layer output \(\widehat{U}_t\)

\For{\(r=0\) to \(R\)}
    \State \(Q_t \leftarrow \mathrm{CriticAgent}(\Omega_t,A_t,U_t)\)
    \State Parse \(Q_t=\{v_t,\ell_t,b_t\}\)

    \If{\(v_t=\mathrm{approve}\) or \(r=R\)}
        \State \Return \(\widehat{U}_t \leftarrow U_t\)
    \EndIf

    \If{\(\ell_t \in \{\mathrm{trigger},\mathrm{both}\}\)}
        \State \(A_t \leftarrow \mathrm{TriggerAnalyzer}(\Omega_t,b_t)\)
    \EndIf

    \State \(U_t \leftarrow \mathrm{AppraisalLayer}(A_t,E_{t-1},\mathcal{Y},b_t)\)
\EndFor
\end{algorithmic}
\end{algorithm}

\section{Experiment Setup}

\subsection{Evaluation Tasks}

We recruited a total of 39 nursing students and evaluated their 39 multi-turn conversations with CPM-MultiAgent alongside 120 controlled single-trigger cases. Both tasks assessed the simulated patient's emotion updates and responses; the conversations also assessed coherence across turns. Task definitions and assessment units are detailed in Appendix~\ref{sec:evaluation_tasks_appendix} (Table~\ref{tab:evaluation_sets}).

\subsection{Dataset}

\textbf{Patient profiles.} Nursing experts constructed 12 synthetic patient profiles based on cases and conditions commonly encountered in nursing communication. Four nursing experts reviewed and confirmed all 12 profiles before use. The profiles were written with neutral emotional framing and served as contextual inputs; differences among profiles were not themselves evaluated.

\textbf{Controlled single-trigger cases.} Nursing experts constructed 120 cases spanning the communication-input categories detailed in Appendix~\ref{sec:evaluation_tasks_appendix}. Each case specified a patient context, an initial emotion state, and one communication input; CPM-MultiAgent generated the patient response and updated emotion state. The nursing experts involved in material construction and review did not participate in the subsequent rating of system outputs.

\subsection{Evaluation Criteria}

Evaluation uses six separately rated dimensions: Emotional Update Plausibility (EUP), Appraisal Reasoning Quality (ARQ), Temporal Consistency (TC), Behavioural Expression Quality (BEQ), Clinical Realism (CR), and Simulation Quality (SQ). Definitions appear in Table~\ref{tab:evaluation_metrics}; TC applies only to complete conversations. These evaluation ratings are distinct from the emotion-intensity scores.

For controlled-case comparisons, ALL denotes the per-case, equal-weight mean of EUP, ARQ, BEQ, CR, and SQ within each judge. SQ remains a separately assigned criterion, whereas ALL summarizes the five applicable ratings. We report $\Delta_{\mathrm{ALL}}$, the mean within-case difference in ALL between CPM-MultiAgent and its comparator; positive values favor CPM-MultiAgent.

\subsection{Evaluation Protocol}

\textbf{Nursing expert evaluation.} Twelve nursing experts with at least three years of clinical experience rated system outputs on a five-point Likert scale. They also judged anonymized baseline pairs on EUP and ARQ. Assignment, scoring, and aggregation procedures are detailed in Appendix~\ref{sec:expert_eval_procedure}.

\textbf{Nursing student evaluation.} After completing their dialogue, students rated five questionnaire items on their interaction experience using the same five-point Likert scale. These experience ratings were separate from the experts' criterion-level output ratings. The procedure and questionnaire statements are provided in Appendix~\ref{sec:student_eval_procedure}.

\textbf{LLM-as-Judge evaluation.} We used four judges from different providers: GPT-6.1 Sol~\citep{openai2026gpt61sol}, Claude Opus 5.5~\citep{anthropic2026opus55}, DeepSeek-V4-Pro~\citep{deepseek2026v4}, and Gemini 3.1 Pro~\citep{google2026gemini31pro}. Each independently scored the same evaluation materials on a 1--5 scale using the task-specific criteria. Ratings were summarized separately by judge as mean $\pm$ standard deviation (SD) with 95\% confidence intervals (CIs). Scoring and bootstrap procedures appear in Appendix~\ref{sec:llm_eval_procedure}.

\subsection{Baseline}

Baseline comparisons use matched patient contexts, initial emotion states, and communication inputs across 120 controlled cases. The six baselines comprise Zero-shot and Few-shot~\citep{brown2020language}, Zero-shot chain-of-thought (CoT)~\citep{kojima2022large}, Few-shot CoT~\citep{wei2022chain}, Self-consistency~\citep{wang2022self}, and Self-Refine~\citep{madaan2023self}.

\subsection{Implementation Details}

CPM-MultiAgent uses LangGraph and OpenAI GPT-5.6 Luna with medium reasoning effort. Luna was selected to balance patient simulation capability against cumulative cost and latency in multi-turn interaction. The backbone comparison additionally includes DeepSeek-V4.1-Flash, Claude Haiku 4.5, Gemini 3.5 Flash-Lite~\citep{google2026gemini35flashlite}, and the open-weight models GPT-OSS 120B and Gemma 4 26B A4B. For evaluation, GPT-6.1 Sol, Claude Opus 5.5, and Gemini 3.1 Pro used medium reasoning effort; DeepSeek-V4-Pro used high. These are provider-specific settings and do not imply equal computational budgets across models. The CPM-MultiAgent, baseline, and judge prompts are provided in Appendix~\ref{sec:prompt_appendix}.

\section{Experiment Results}

\begin{table*}[t]
\centering
\small
\setlength{\tabcolsep}{6pt}
\begin{tabular}{@{}lcccc@{}}
\toprule
& \multicolumn{2}{c}{\textbf{Student-led dialogue ($n=39$)}}
& \multicolumn{2}{c}{\textbf{Controlled trigger ($n=120$)}} \\
\cmidrule(lr){2-3}\cmidrule(l){4-5}
\textbf{Criterion}
& \textbf{Mean $\pm$ SD}
& \textbf{95\% CI}
& \textbf{Mean $\pm$ SD}
& \textbf{95\% CI} \\
\midrule
EUP & $4.41 \pm 0.35$ & [4.31, 4.52] & $4.55 \pm 0.33$ & [4.49, 4.61] \\
ARQ & $4.43 \pm 0.39$ & [4.30, 4.55] & $4.50 \pm 0.34$ & [4.44, 4.56] \\
TC & $4.48 \pm 0.35$ & [4.37, 4.59] & -- & -- \\
BEQ & $4.45 \pm 0.32$ & [4.35, 4.54] & $4.53 \pm 0.33$ & [4.46, 4.58] \\
CR & $3.95 \pm 0.34$ & [3.85, 4.05] & $3.98 \pm 0.26$ & [3.94, 4.03] \\
SQ & $4.22 \pm 0.34$ & [4.11, 4.32] & $4.23 \pm 0.31$ & [4.18, 4.28] \\
\bottomrule
\end{tabular}
\caption{Nursing-expert ratings of CPM-MultiAgent on 39 conversations and 120 controlled cases. Entries are mean $\pm$ sample SD of case-level expert means and 95\% percentile-bootstrap CIs (10,000 resamples). A dash indicates inapplicable TC.}
\label{tab:expert_eval_results}
\end{table*}

\begin{table*}[t]
\centering
\footnotesize
\setlength{\tabcolsep}{2pt}
\begin{tabular}{@{}l*{4}{>{\centering\arraybackslash}p{0.112\textwidth}>{\centering\arraybackslash}p{0.098\textwidth}}@{}}
\toprule
\multirow{2}{*}{\textbf{Criterion}}
& \multicolumn{2}{c}{\textbf{GPT-6.1 Sol}}
& \multicolumn{2}{c}{\textbf{Claude Opus 5.5}}
& \multicolumn{2}{c}{\textbf{DeepSeek-V4-Pro}}
& \multicolumn{2}{c}{\textbf{Gemini 3.1 Pro}} \\
\cmidrule(lr){2-3}
\cmidrule(lr){4-5}
\cmidrule(lr){6-7}
\cmidrule(l){8-9}
& \textbf{Mean $\pm$ SD} & \textbf{95\% CI}
& \textbf{Mean $\pm$ SD} & \textbf{95\% CI}
& \textbf{Mean $\pm$ SD} & \textbf{95\% CI}
& \textbf{Mean $\pm$ SD} & \textbf{95\% CI} \\
\midrule

\multicolumn{9}{l}{\textit{(a) Student-led dialogue ($n=39$)}} \\
\addlinespace[2pt]

EUP
& $3.10 \pm 0.55$ & \mbox{[2.92, 3.28]}
& $2.69 \pm 0.77$ & \mbox{[2.46, 2.92]}
& $3.46 \pm 0.55$ & \mbox{[3.28, 3.64]}
& $3.10 \pm 0.55$ & \mbox{[2.95, 3.28]} \\

ARQ
& $3.03 \pm 0.54$ & \mbox{[2.85, 3.18]}
& $2.59 \pm 0.68$ & \mbox{[2.38, 2.79]}
& $3.13 \pm 0.70$ & \mbox{[2.90, 3.33]}
& $3.05 \pm 0.51$ & \mbox{[2.90, 3.21]} \\

TC
& $3.54 \pm 0.68$ & \mbox{[3.33, 3.74]}
& $3.56 \pm 0.75$ & \mbox{[3.33, 3.79]}
& $3.90 \pm 0.72$ & \mbox{[3.67, 4.13]}
& $3.18 \pm 0.60$ & \mbox{[3.00, 3.36]} \\

BEQ
& $3.85 \pm 0.63$ & \mbox{[3.64, 4.03]}
& $3.49 \pm 0.68$ & \mbox{[3.28, 3.69]}
& $4.03 \pm 0.71$ & \mbox{[3.82, 4.23]}
& $3.67 \pm 0.48$ & \mbox{[3.51, 3.79]} \\

CR
& $3.15 \pm 0.59$ & \mbox{[2.97, 3.33]}
& $2.90 \pm 0.75$ & \mbox{[2.67, 3.13]}
& $3.44 \pm 0.60$ & \mbox{[3.26, 3.62]}
& $3.13 \pm 0.52$ & \mbox{[2.97, 3.28]} \\

\textbf{SQ}
& $3.21 \pm 0.57$ & \mbox{[3.03, 3.38]}
& $2.85 \pm 0.71$ & \mbox{[2.64, 3.08]}
& $3.41 \pm 0.55$ & \mbox{[3.26, 3.59]}
& $3.15 \pm 0.54$ & \mbox{[2.97, 3.31]} \\

\midrule
\multicolumn{9}{l}{\textit{(b) Controlled single-trigger cases ($n=120$)}} \\
\addlinespace[2pt]

EUP
& $4.91 \pm 0.32$ & \mbox{[4.85, 4.96]}
& $4.23 \pm 0.48$ & \mbox{[4.15, 4.32]}
& $4.81 \pm 0.42$ & \mbox{[4.73, 4.88]}
& $3.89 \pm 0.38$ & \mbox{[3.83, 3.96]} \\

ARQ
& $4.92 \pm 0.28$ & \mbox{[4.87, 4.96]}
& $3.97 \pm 0.48$ & \mbox{[3.88, 4.05]}
& $4.60 \pm 0.53$ & \mbox{[4.51, 4.69]}
& $3.63 \pm 0.52$ & \mbox{[3.54, 3.73]} \\

TC
& -- & --
& -- & --
& -- & --
& -- & -- \\

BEQ
& $4.58 \pm 0.62$ & \mbox{[4.46, 4.68]}
& $3.76 \pm 0.47$ & \mbox{[3.68, 3.84]}
& $4.67 \pm 0.61$ & \mbox{[4.55, 4.78]}
& $3.59 \pm 0.53$ & \mbox{[3.50, 3.68]} \\

CR
& $4.87 \pm 0.37$ & \mbox{[4.80, 4.93]}
& $3.99 \pm 0.33$ & \mbox{[3.93, 4.05]}
& $4.76 \pm 0.50$ & \mbox{[4.67, 4.84]}
& $3.88 \pm 0.36$ & \mbox{[3.81, 3.93]} \\

\textbf{SQ}
& $4.79 \pm 0.43$ & \mbox{[4.72, 4.87]}
& $3.94 \pm 0.33$ & \mbox{[3.88, 4.00]}
& $4.68 \pm 0.52$ & \mbox{[4.58, 4.77]}
& $3.74 \pm 0.44$ & \mbox{[3.66, 3.82]} \\

\addlinespace[2pt]
ALL
& $4.81 \pm 0.33$ & \mbox{[4.75, 4.87]}
& $3.98 \pm 0.29$ & \mbox{[3.93, 4.03]}
& $4.70 \pm 0.36$ & \mbox{[4.64, 4.76]}
& $3.75 \pm 0.33$ & \mbox{[3.69, 3.80]} \\

\bottomrule
\end{tabular}

\caption{LLM-judge ratings of CPM-MultiAgent on (a) 39 conversations and (b) 120 controlled cases. Entries are mean $\pm$ sample SD and 95\% percentile-bootstrap CIs (10,000 resamples), computed separately by task and judge. In (b), ALL averages the five applicable criteria within each case. A dash indicates inapplicable TC.}
\label{tab:llm_primary_eval_results}
\end{table*}

\begin{table*}[t]
\centering
\small
\setlength{\tabcolsep}{5pt}
\begin{tabular}{@{}lcccc@{}}
\toprule
\multirow{2}{*}{\textbf{CPM-MultiAgent vs.}}
& \multicolumn{2}{c}{\textbf{EUP}}
& \multicolumn{2}{c}{\textbf{ARQ}} \\
\cmidrule(lr){2-3}\cmidrule(lr){4-5}
& W/L/T & Non-tied win rate
& W/L/T & Non-tied win rate \\
\midrule
Zero-shot
& 37/6/7  & 86.05\%
& 39/5/6  & 88.64\% \\
Few-shot
& 35/7/8  & 83.33\%
& 36/7/7  & 83.72\% \\
Zero-shot CoT
& 34/8/8  & 80.95\%
& 35/8/7  & 81.40\% \\
Few-shot CoT
& 32/9/9  & 78.05\%
& 33/9/8  & 78.57\% \\
Self-consistency
& 33/8/9  & 80.49\%
& 34/8/8  & 80.95\% \\
Self-Refine
& 31/10/9 & 75.61\%
& 32/10/8 & 76.19\% \\
\bottomrule
\end{tabular}
\caption{Descriptive nursing-expert preferences on 50 anonymized pairs per baseline. W/L/T denote judgments favoring CPM-MultiAgent, favoring the baseline, and ties. Non-tied win rate is $W/(W+L)$.}
\label{tab:human_eval_results}
\end{table*}

\begin{table}[tbp]
\centering
\small
\setlength{\tabcolsep}{3pt}
\begin{tabular*}{\columnwidth}{@{}l@{\extracolsep{\fill}}cr@{}}
\toprule
\textbf{Item}
& \textbf{Mean}
& \shortstack[r]{\textbf{Ratings 4--5}\\\textbf{$n/N$ (\%)}} \\
\midrule
Contextual appropriateness & 4.15 & 34/39 (87.18\%) \\
Emotional realism         & 4.21 & 32/39 (82.05\%) \\
Communication insight     & 4.05 & 34/39 (87.18\%) \\
Practice value    & 4.31 & 34/39 (87.18\%) \\
Satisfaction and reuse           & 4.33 & 34/39 (87.18\%) \\
\bottomrule
\end{tabular*}
\caption{Descriptive student feedback ($N=39$). Entries show mean ratings (1--5) and the count/proportion of responses rated 4--5.}
\label{tab:student_experience_results}
\end{table}

\subsection{Expert Evaluation}
\label{sec:expert_evaluation}

Nursing experts rated CPM-MultiAgent's student-led conversations at 4.22 for SQ and 4.48 for TC (Table~\ref{tab:expert_eval_results}). Controlled single-trigger cases received a 4.55 EUP score for patient-specific emotion updates. These ratings assess complementary aspects of emotion simulation; they do not establish differences between tasks.

\subsection{LLM-as-Judge Evaluation}
\label{sec:llm_judge_evaluation}

The controlled single-trigger cases received favorable ratings across all four judges (Table~\ref{tab:llm_primary_eval_results}). EUP means ranged from 3.89 to 4.91, and SQ means ranged from 3.74 to 4.79 on the 1--5 scale. GPT-6.1 Sol and DeepSeek-V4-Pro rated all five applicable criteria above 4, supporting plausible emotion updates and credible patient simulation under controlled inputs.

In student-led dialogues, BEQ and TC were the two highest-rated dimensions for every judge, with respective mean ranges of 3.49--4.03 and 3.18--3.90. These ratings identify emotional expression and continuity as relative strengths during multi-turn interaction. SQ means were more moderate (2.85--3.41), indicating that overall simulation quality remains a challenge in naturally unfolding conversations.

Expert ratings provided complementary support, with SQ means of 4.23 for controlled cases and 4.22 for complete dialogues (Table~\ref{tab:expert_eval_results}). Experts rated every dialogue criterion higher than all four LLM judges.

\subsection{Baseline Comparison}

Nursing experts preferred CPM-MultiAgent over all six baselines on both EUP and ARQ (Table~\ref{tab:human_eval_results}). Non-tied win rates ranged from 75.61\% to 86.05\% for EUP and from 76.19\% to 88.64\% for ARQ. Preferences remained favorable against reasoning and refinement baselines, supporting the perceived plausibility of the emotion updates and their explanations. These preference results are descriptive, with ties reported separately.

On the same 120 controlled cases, CPM-MultiAgent achieved higher ALL means than every baseline under all four judges (Appendix~\ref{sec:baseline_appendix}, Table~\ref{tab:llm_judge_results}). Gains over Zero-shot, Few-shot, and Zero-shot CoT ranged from 0.20 to 0.73 points, with all Holm-adjusted $p<0.001$. Improvements over stronger comparators were smaller (0.07--0.24 points) and not uniformly significant, although the direction remained favorable across judges. Criterion-level results also showed higher scores than Zero-shot and Few-shot on all five criteria under every judge (Holm-adjusted $p<0.05$). The complete criterion-level comparisons appear in Appendix~\ref{sec:baseline_appendix} (Tables~\ref{tab:baseline_gpt61sol}--\ref{tab:baseline_deepseekv4pro}).

\subsection{Ablation Study}

The full system achieved higher ALL means than all seven component-removal variants under every judge (Appendix~\ref{sec:ablation_appendix}, Table~\ref{tab:ablation_results}). Removing Trigger Analysis or Implication Check reduced ALL by 0.14--0.37 and 0.11--0.37 points, respectively, with all Holm-adjusted $p<0.05$. These results support the contribution of patient-specific trigger grounding and consequence appraisal to single-trigger simulation quality.

Criterion-level results showed that removing Implication Check reduced ARQ under all four judges (Holm-adjusted $p<0.001$), directly supporting its role in appraisal explanations. Cross-Appraisal Peer Review also improved ALL under every judge (Holm-adjusted $p<0.05$), supporting refinement across appraisal perspectives. Normative Significance Check and Consistency Critique yielded smaller, judge-dependent gains, with non-significant ALL differences under DeepSeek-V4-Pro. Detailed criterion-level results appear in Appendix~\ref{sec:ablation_appendix} (Tables~\ref{tab:ablation_gpt61sol}--\ref{tab:ablation_deepseekv4pro}); these comparisons do not assess multi-turn TC.

\subsection{Backbone Robustness Study}
\label{sec:backbone_robustness}

We evaluated six generation backbones using the same framework, 120 controlled cases, and four fixed judges. Across GPT-5.6 Luna and the three alternative proprietary backbones, within-judge SQ mean ranges spanned 0.06--0.26 points. Both open-weight backbones received lower SQ means than all four proprietary backbones under every judge. These descriptive results support use across the tested proprietary backbones while indicating sensitivity to backbone selection; they do not establish equivalence. Criterion-level means, SDs, and bootstrap CIs appear in Appendix~\ref{sec:backbone_appendix} (Tables~\ref{tab:backbone_gpt61sol}--\ref{tab:backbone_deepseekv4pro}).

\subsection{Student User Study}

Students reported positive experiences across all five questionnaire items (Table~\ref{tab:student_experience_results}); full statements appear in Table~\ref{tab:student_questionnaire}. Mean scores ranged from 4.05 to 4.33 out of 5, and 82.05--87.18\% of responses were ratings of 4 or 5. Satisfaction and reuse received the highest mean score (4.33), followed by practice value (4.31). Students valued its contextual appropriateness, emotional realism, and practice utility.

\section{Conclusion}

CPM-MultiAgent models patient emotion updates through trigger analysis, structured appraisal, state integration, and consistency critique. Expert preferences and matched comparisons support improved controlled-case simulation, with consistent contributions from trigger analysis and implication appraisal. Student-led dialogues highlight emotional expression and continuity as relative strengths, alongside scope for improving multi-turn simulation quality.

\section*{Limitations}

CPM-MultiAgent requires multiple inference steps to update patient emotions, which may introduce response latency during multi-turn dialogue. Deployment therefore needs to consider the trade-off between appraisal quality, response speed, and the timeliness of emotional updates. Future work will examine how to reduce inference overhead while preserving coherent emotion updates and responsive interaction.

\section*{Ethical Considerations}

The human evaluation in this study was approved by the institutional ethics review board. We recruited a total of 39 currently enrolled nursing students for dialogue interactions and 12 nursing experts for output evaluation. The nursing experts involved in preparing and reviewing the synthetic evaluation materials were distinct from those recruited to rate system outputs. All participants took part voluntarily and anonymously. Before the evaluation, participants were informed of the purpose of the study, the evaluation procedure, and their right to withdraw at any time. The collected responses were used only for research purposes and were analyzed in aggregate. No personally identifiable information was collected or reported. Potential risks of this work include the possibility that emotionally adaptive dialogue agents may produce overly persuasive, misleading, or inappropriate emotional responses, especially in sensitive settings such as healthcare, education, or counseling simulation. The proposed framework is intended for controlled role-simulation and training scenarios rather than autonomous clinical, educational, or counseling decision-making. Human oversight is necessary when deploying such systems in high-stakes domains, and outputs should not be treated as professional advice. We manually checked the constructed dialogue trials and evaluation materials to ensure that they do not contain personally identifying information or offensive content. Patient profiles and controlled single-trigger cases were synthetically constructed, whereas student-led conversations were collected for this evaluation. We use third-party models, frameworks, and referenced artifacts in accordance with their stated licenses or terms of use, and we do not redistribute restricted artifacts.

\bibliography{custom}

\appendix

\section{CPM Appraisal Checks}
\label{sec:cpm_appendix}

\begin{table*}[!t]
  \centering
  \small
  \setlength{\tabcolsep}{6pt}
  \renewcommand{\arraystretch}{1.35}
  \setlength{\extrarowheight}{2pt}

  \begin{tabular}{|
    >{\raggedright\arraybackslash}m{0.17\textwidth}|
    >{\raggedright\arraybackslash}m{0.47\textwidth}|
    >{\raggedright\arraybackslash}m{0.24\textwidth}|
  }
    \hline
    \textbf{Appraisal Check} &
    \textbf{Information Assessed during Appraisal} &
    \textbf{SECs} \\
    \hline

    \textbf{Relevance} &
    How relevant is this event for me? \newline
    Does it directly affect me or my social group? &
    Novelty \newline
    Pleasantness \newline
    Goal Relevance \\
    \hline

    \textbf{Implications} &
    What are the implications or consequences of this event? \newline
    How do these implications affect my well-being? \newline
    How do these implications affect my short- or long-term goals? &
    Causal Attribution \newline
    Possible Outcomes \newline
    Discrepancy from Expectation \newline
    Goal Conduciveness \newline
    Urgency \\
    \hline

    \textbf{Coping Potential} &
    How well can I cope with or adjust to these consequences? &
    Controllability \newline
    Power/Resources \newline
    Coping Potential \\
    \hline

    \rule[-0.8em]{0pt}{4.8em}\shortstack[l]{\textbf{Normative}\\\textbf{Significance}\\\textbf{Evaluation}} &
    What is the significance of this event with respect to my self-concept and social norms and values? &
    Internal Standards Check \newline
    External Standards Check \\
    \hline
  \end{tabular}

  \caption{CPM appraisal checks and corresponding SECs.}
  \label{tab:cpm}
\end{table*}

\section{Patient Emotion Criteria}
\label{sec:patient_emotion_criteria}
Table~\ref{tab:patient_emotion_criteria} lists the nursing communication criteria developed through discussions with nursing experts. The eight original categories were retained from the source study identified in Section~\ref{sec:patient_emotion_representation}; \textit{nervous} and \textit{inspired} are additions made for our nursing communication setting. All criteria describe how we interpret the categories in this setting, rather than definitions reproduced from the source study. They identify patient-specific cues relevant to appraisal and emotion updating.

\begin{table*}[!t]
\centering
\small
\setlength{\tabcolsep}{5pt}
\renewcommand{\arraystretch}{1.22}
\begin{tabular}{@{}>{\raggedright\arraybackslash}p{0.10\textwidth}>{\raggedright\arraybackslash}p{0.14\textwidth}>{\raggedright\arraybackslash}p{0.69\textwidth}@{}}
\toprule
\textbf{Valence} & \textbf{Emotion} & \textbf{Criterion in nurse--patient communication} \\
\midrule
Positive & Relief & Whether burdens, worries, uncertainty, threats, pain, or specific concerns have been alleviated. \\
Positive & Satisfaction & Whether the patient's needs, information needs, concerns, or decision-making goals have been met or substantively addressed. \\
Positive & Joy & Whether a clear positive outcome with strong positive emotional significance has occurred. \\
Positive & Interest & Whether meaningful engagement with information relevant to the patient has increased. \\
Positive & Inspired & Whether hope, agency, confidence, or motivation has increased. \\
\midrule
Negative & Anger & Whether obstruction, unfairness, blame, disrespect, or a violation of norms has occurred. \\
Negative & Fear & Whether an identifiable threat or danger that the patient perceives as credible has become salient. \\
Negative & Regret & Whether self-evaluation, attribution of personal responsibility, or ``if only'' counterfactual thinking has become salient. \\
Negative & Sadness & Whether a sense of loss, helplessness, or irreversibility has increased. \\
Negative & Nervous & Whether uncertainty, waiting, unpredictability, lack of clarity, interpersonal pressure, or unconfirmed adverse possibilities have increased. \\
\bottomrule
\end{tabular}
\caption{Operational criteria for ten patient emotion categories in nursing communication, including the added categories inspired and nervous.}
\label{tab:patient_emotion_criteria}
\end{table*}

\section{Scenario and Emotion Settings}
\label{sec:scenario_appendix}

\subsection{Evaluation Task Definitions}
\label{sec:evaluation_tasks_appendix}

The two evaluation tasks assess complementary aspects of patient emotion simulation, as summarized in Table~\ref{tab:evaluation_sets}. Student-led conversations assess patient responses and emotional trajectories during naturally unfolding nursing communication. Controlled single-trigger cases assess an individual response and emotion transition given a specified patient context, initial state, and communication input. Both tasks assess the simulated patient; they do not measure the student's communication competence.

\begin{table}[tbp]
\centering
\small
\begin{tabular}{@{}lcp{0.43\columnwidth}@{}}
\toprule
\textbf{Task} & \textbf{$n$} & \textbf{Unit assessed} \\
\midrule
Student-led dialogue & 39 & Complete conversation and turn-by-turn emotion trajectory \\
Controlled trigger & 120 & One context-dependent input, response, and emotion transition \\
\bottomrule
\end{tabular}
\caption{Complementary evaluation tasks. TC is assessed only for complete student-led conversations.}
\label{tab:evaluation_sets}
\end{table}

\textbf{Student-led nursing communication.} Complete conversations allow assessment of whether emotion changes remain plausible in the patient's clinical and conversational context and coherent across turns. The unit assessed is the entire conversation, including its sequence of patient emotion updates.

\textbf{Controlled single-trigger evaluation.} Controlled cases complement the typically appropriate and reassuring inputs produced by nursing students with positive, negative, neutral, mixed, and less common or more extreme inputs. Each case assesses a context-dependent emotion transition and generated response; continuity across a conversation is not assessed.

\subsection{Emotion Taxonomies and Intensity}

For nursing communication, we use the ten patient emotion categories defined in Section~\ref{sec:patient_emotion_representation}, with operational criteria in Appendix~\ref{sec:patient_emotion_criteria}. The framework accepts a predefined label set as input and can accommodate alternative emotion taxonomies for other simulation settings. Each label receives a separate intensity score, allowing multiple emotions to coexist. We use the five intensity anchors from the Positive and Negative Affect Schedule (PANAS)~\citep{watson1988development}, shown in Table~\ref{tab:emotion_intensity}, to score each category in our adapted representation.

\begin{table*}[!t]
\centering
\small
{\renewcommand{\arraystretch}{1.55}
\begin{tabular}{|c|p{3.5cm}|}
\hline
\textbf{Intensity} & \textbf{Description} \\
\hline
1 & Very slightly or not at all. \\
\hline
2 & A little. \\
\hline
3 & Moderately. \\
\hline
4 & Quite a bit. \\
\hline
5 & Extremely. \\
\hline
\end{tabular}
}
\caption{Emotion intensity scale used for each emotion label.}
\label{tab:emotion_intensity}
\end{table*}

\section{Evaluation Metrics}
\label{sec:metric_appendix}

SQ is a separate holistic judgment rather than an average of the other dimensions. TC is assessed only for complete student-led conversations.

\begin{table*}[!t]
\centering
\small
{\renewcommand{\arraystretch}{1.55}
\begin{tabular}{|p{1.6cm}|p{11.4cm}|}
\hline
\textbf{Metric} & \textbf{Description} \\
\hline
EUP
& Considering the patient's previous emotional state, the current interaction, patient characteristics, and clinical context, whether the direction and magnitude of the emotional change, as well as the resulting emotional state, are plausible. \\
\hline

ARQ
& Whether the explanation for the patient's emotional change is plausible, including whether it is grounded in relevant events, information, or communication in the current interaction, reflects a reasonable appraisal of these triggers, and explains the resulting emotional change. \\
\hline

TC
& Whether the patient's current emotional state and its transition remain coherent with the previous emotional state and prior interactions, without unexplained abrupt changes, contradictions, or emotional resets. \\
\hline

BEQ
& Whether the patient's language and behavioural expressions appropriately reflect their current emotional state. \\
\hline

CR
& Whether the intensity, progression, and expression of the patient's emotional responses are consistent with what could plausibly occur in real nurse--patient communication. \\
\hline

SQ
& Whether the patient simulation provides a credible, coherent, and clinically appropriate dynamic emotional interaction experience. \\
\hline
\end{tabular}
}
\caption{Evaluation criteria used in the quantitative and qualitative evaluation.}
\label{tab:evaluation_metrics}
\end{table*}

\subsection{LLM-as-Judge Evaluation Procedure}
\label{sec:llm_eval_procedure}

Each judge independently rated the same patient contexts and generated outputs using the task-specific criteria. In the primary evaluation, each judge assessed every complete conversation and controlled case once. All applicable ratings were valid, with none missing or excluded. Each candidate in controlled-case comparisons was assessed separately. The shared system prompt and task-specific instructions appear in Appendix~\ref{sec:prompt_appendix}.

Ratings were summarized separately by task and judge as the mean and sample SD across assessed units. We estimated 95\% CIs using 10,000 percentile bootstrap resamples. Each resample drew complete conversations or case records with replacement within the corresponding task and judge. The 2.5th and 97.5th percentiles of bootstrap means formed the CI bounds. The same case-level procedure summarized the backbone ratings. Baseline and ablation comparisons used the paired tests and separate Holm correction families specified in Appendices~\ref{sec:baseline_appendix} and~\ref{sec:ablation_appendix}.

ALL comparisons were computed in Python 3.11.14 using NumPy 2.3.2 and SciPy 1.17.1, with all 120 matched cases retained ($df=119$). Controlled-case criterion summaries and within-judge ALL summaries used bootstrap seed 20261013.

\begingroup

\Needspace{7\baselineskip}
\begin{prompttext}
\section{Prompts}
\label{sec:prompt_appendix}

This section provides the CPM-MultiAgent, baseline, and LLM-as-Judge prompts used in our experiments. Structured outputs use JavaScript Object Notation (JSON).

\subsection{CPM-MultiAgent Prompts}

\subsubsection{Trigger Analyzer Agent Prompt}
\end{prompttext}
\begin{promptbox}{Trigger Analyzer Agent Prompt}

\textbf{Role Description:} You are the Trigger Analyzer in a publication-grade multi-agent CPM emotion-update pipeline. Your task is to identify the affectively salient dialogue trigger before any emotion appraisal is performed.

\textbf{Core Objective:} Convert the latest user input into a structured trigger analysis pack. The analysis should specify what happened, how the trigger is grounded in the persona and dialogue context, and which CPM-relevant cues should guide downstream appraisal agents.

\textbf{Your Goals:}
\begin{enumerate}
    \item Identify the objective trigger structure, including the event, participants, action, dialogue position, and relevant span in the latest user input.
    \item Connect the trigger to the persona's identity, goals, relationships, dialogue memory, and previous emotional state.
    \item Extract CPM-relevant cues for relevance, implication, coping potential, and normative significance, without performing the full appraisal.
\end{enumerate}

\textbf{Input:}
\begin{itemize}
    \item Persona profile and scenario context: \texttt{\{knowledge\}}
    \item Dialogue memory: \texttt{\{memory\}}
    \item Recent dialogue history: \texttt{\{recent\_history\}}
    \item Previous emotion state: \texttt{\{last\_emotions\}}
    \item Latest user input: \texttt{"\{user\_input\}"}
    \item Optional critic feedback: \texttt{\{critic\_feedback\}}
\end{itemize}

\textbf{Constraints:}
\begin{itemize}
    \item Do not infer specific emotions or assign emotion intensities.
    \item Do not perform full relevance, implication, coping potential, or normative significance appraisal.
    \item Do not update the emotional state.
    \item Do not introduce information that is not grounded in the given context.
    \item Provide concise evidence-based justification rather than hidden reasoning steps.
\end{itemize}

\textbf{Output Format:}
\begin{lstlisting}[style=jsonstyle]
{
  "trigger_summary": "Two to four sentences summarizing the trigger landscape.",
  "primary_trigger": "The main trigger in the latest user input.",
  "trigger_valence": "Positive | Negative | Mixed | Ambiguous",
  "urgency": "High | Medium | Low",
  "temporal_orientation": "Immediate | Near-term | Long-term | Mixed",
  "social_signal": "Supportive | Neutral | Critical | Hostile | Mixed",
  "controllability_signal": "High | Medium | Low | Mixed | Unclear",
  "target_of_concern": [
    "Affected goal, stake, or entity"
  ],
  "secondary_triggers": [
    "Optional secondary trigger"
  ],
  "appraisal_focus": [
    "Directive for downstream appraisal agents"
  ],
  "evidence": [
    "Evidence point grounded in the input or context"
  ],
  "unresolved_ambiguities": [
    "Uncertainty that should not be over-interpreted"
  ],
  "confidence": "High | Medium | Low",
  "objective_trigger_structure": {
    "event_summary": "Concise description of the objective event.",
    "participants": "People or entities involved.",
    "action": "Central action or state change.",
    "dialogue_position": "Where the trigger appears in the current turn.",
    "trigger_span": "Shortest relevant span from the latest user input."
  },
  "persona_grounded_context": {
    "identity_links": "Connection to the persona's identity.",
    "goal_links": "Connection to the persona's goals or concerns.",
    "relationship_links": "Connection to interpersonal relations.",
    "memory_links": "Connection to prior dialogue memory.",
    "previous_emotion_links": "Connection to the previous emotional state."
  },
  "sec_cues": {
    "relevance": "Cues for novelty, intrinsic pleasantness, and goal relevance.",
    "implication": "Cues for causal attribution, outcome probability, expectation discrepancy, goal conduciveness, and urgency.",
    "coping_potential": "Cues for control, power, and adjustment.",
    "normative_significance": "Cues for internal and external standards."
  }
}
\end{lstlisting}

\textbf{Instruction:} Build the trigger analysis pack using the persona profile, scenario context, dialogue memory, recent dialogue history, previous emotion state, and latest user input. If critic feedback is provided, revise the analysis while preserving the same output schema.

\end{promptbox}
\Needspace{7\baselineskip}
\begin{prompttext}
\subsubsection{Relevance Appraisal Agent Prompt}
\end{prompttext}
\begin{promptbox}{Relevance Appraisal Agent Prompt}

\textbf{Role Description:} You are the Relevance Appraisal Agent in a CPM-based emotional evaluation pipeline. Your task is to determine whether the latest user input is relevant enough to trigger further emotional processing for the simulated persona.

\textbf{Appraisal Goal:} Perform only the CPM Relevance Check. Evaluate whether the trigger matters to the persona's goals, well-being, relationships, or current stakes. Do not infer specific emotions or update the emotional state.

\textbf{Appraisal Dimensions:}
\begin{enumerate}
    \item \textbf{Novelty:} Assess whether the input is new, unexpected, surprising, or predictable for the persona.
    \item \textbf{Intrinsic Pleasantness:} Assess whether the event or information is inherently pleasant, unpleasant, mixed, or neutral.
    \item \textbf{Goal Relevance:} Assess whether the input affects the persona's goals, needs, relationships, status, or well-being.
\end{enumerate}

\textbf{Input:}
\begin{itemize}
    \item Persona profile and scenario context: \{knowledge\}
    \item Dialogue memory: \{memory\}
    \item Trigger analysis pack: \{trigger\_analysis\}
    \item Latest user input: ``\{user\_input\}''
    \item Optional critic feedback: \{critic\_feedback\}
\end{itemize}

\textbf{Constraints:}
\begin{itemize}
    \item Do not assign emotion labels such as fear, anger, sadness, or joy.
    \item Do not evaluate implication, coping potential, or normative significance.
    \item Do not introduce information that is not grounded in the provided context.
    \item Provide concise evidence-based justification rather than hidden reasoning steps.
\end{itemize}

\textbf{Output Format:}
\begin{lstlisting}[style=jsonstyle]
{
  "relevance": "High | Medium | Low",
  "content": "Two to three sentences explaining how relevant the input is to the persona's goals, well-being, relationships, or current stakes.",
  "evidence": [
    "Evidence point grounded in the persona context, trigger analysis, or latest input"
  ],
  "confidence": "High | Medium | Low",
  "possible_concern": "A short caveat, uncertainty, or condition that could weaken this judgement.",
  "sec_assessments": {
    "novelty": "Short assessment of whether the input is new, unexpected, surprising, or predictable.",
    "intrinsic_pleasantness": "Short assessment of whether the input is pleasant, unpleasant, mixed, or neutral.",
    "goal_relevance": "Short assessment of how strongly the input affects the persona's goals, needs, relationships, status, or well-being."
  },
  "upstream_appraisal_influence": ""
}
\end{lstlisting}

\textbf{Instruction:} Perform the Relevance Check using the trigger analysis pack, persona context, dialogue memory, and latest user input. If critic feedback is provided, revise the appraisal while keeping the same output schema.

\end{promptbox}
\Needspace{7\baselineskip}
\begin{prompttext}
\subsubsection{Implication Appraisal Agent Prompt}
\end{prompttext}
\begin{promptbox}{Implication Appraisal Agent Prompt}

\textbf{Role Description:} You are the Implication Appraisal Agent in a CPM-based emotional evaluation pipeline. Your task is to evaluate the potential consequences of the latest user input for the simulated persona.

\textbf{Core Objective:} Perform only the CPM Implication Check. Assess how the trigger may affect the persona's future state, goals, well-being, relationships, and current stakes, while taking the prior relevance appraisal as given.

\textbf{Your Goals:}
\begin{enumerate}
    \item Identify the likely cause of the situation, such as the self, another person, or external circumstances.
    \item Assess the likely short-term or long-term consequences for the persona.
    \item Determine whether the trigger matches or violates the persona's expectations.
    \item Evaluate whether the trigger supports or obstructs the persona's goals and needs.
    \item Judge whether the situation requires an immediate response.
\end{enumerate}

\textbf{Input:}
\begin{itemize}
    \item Persona profile and scenario context: \texttt{\{knowledge\}}
    \item Dialogue memory: \texttt{\{memory\}}
    \item Trigger analysis pack: \texttt{\{trigger\_analysis\}}
    \item Relevance appraisal: \texttt{\{relevance\_appraisal\}}
    \item Latest user input: \texttt{"\{user\_input\}"}
    \item Optional critic feedback: \texttt{\{critic\_feedback\}}
\end{itemize}

\textbf{Constraints:}
\begin{itemize}
    \item Do not re-evaluate relevance; treat the relevance appraisal as prior input.
    \item Do not evaluate coping potential, including control, power, or adaptability.
    \item Do not update the emotional state.
    \item Do not introduce information that is not grounded in the provided context.
    \item Provide concise evidence-based justification rather than hidden reasoning steps.
\end{itemize}

\textbf{Output Format:}
\begin{lstlisting}[style=jsonstyle]
{
  "classification": "Positive | Negative | Mixed | Uncertain",
  "content": "Two to three sentences explaining how the input affects the persona's future state, objectives, or current stakes.",
  "evidence": [
    "Evidence point grounded in the persona context, trigger analysis, relevance appraisal, or latest input"
  ],
  "confidence": "High | Medium | Low",
  "possible_concern": "A short caveat, uncertainty, or condition that could weaken this judgement.",
  "sec_assessments": {
    "causal_attribution": "Self | Other | Circumstance | Mixed | Unclear",
    "outcome_probability": "Short assessment of likely consequences.",
    "expectation_discrepancy": "Short assessment of whether the trigger matches or violates expectations.",
    "goal_conduciveness": "Conducive | Obstructive | Mixed | Unclear",
    "urgency": "High | Medium | Low"
  },
  "upstream_appraisal_influence": "How the prior relevance appraisal conditions this implication judgement."
}
\end{lstlisting}

\textbf{Instruction:} Perform the Implication Check using the trigger analysis pack, persona context, dialogue memory, relevance appraisal, and latest user input. If critic feedback is provided, revise the appraisal while keeping the same output schema.

\end{promptbox}
\Needspace{7\baselineskip}
\begin{prompttext}
\subsubsection{Coping Potential Appraisal Agent Prompt}
\end{prompttext}
\begin{promptbox}{Coping Potential Appraisal Agent Prompt}

\textbf{Role Description:} You are the Coping Potential Appraisal Agent in a CPM-based emotional evaluation pipeline. Your task is to assess how well the simulated persona can deal with the current situation.

\textbf{Core Objective:} Perform only the CPM Coping Potential Check. Evaluate whether the persona has sufficient control, resources, or adaptability to handle the trigger, while taking the prior relevance and implication appraisals as given.

\textbf{Your Goals:}
\begin{enumerate}
    \item Assess whether the persona can influence or change the event.
    \item Evaluate whether the persona has the knowledge, energy, authority, or social and practical resources needed to respond.
    \item Determine whether the persona can mentally, emotionally, or practically adapt if the event cannot be changed.
\end{enumerate}

\textbf{Scoring Guide:}
\begin{itemize}
    \item \textbf{High:} The persona appears capable of changing the outcome or handling the situation effectively.
    \item \textbf{Medium:} The persona has partial influence or some adaptive resources, but the outcome remains uncertain.
    \item \textbf{Low:} The persona appears helpless, overwhelmed, or lacking the resources needed to act or adapt.
\end{itemize}

\textbf{Input:}
\begin{itemize}
    \item Persona profile and scenario context: \texttt{\{knowledge\}}
    \item Dialogue memory: \texttt{\{memory\}}
    \item Trigger analysis pack: \texttt{\{trigger\_analysis\}}
    \item Relevance appraisal: \texttt{\{relevance\_appraisal\}}
    \item Implication appraisal: \texttt{\{implications\_appraisal\}}
    \item Latest user input: \texttt{"\{user\_input\}"}
    \item Optional critic feedback: \texttt{\{critic\_feedback\}}
\end{itemize}

\textbf{Constraints:}
\begin{itemize}
    \item Do not infer specific emotions or assign emotion intensities.
    \item Do not re-evaluate relevance or implication.
    \item Do not evaluate normative significance.
    \item Do not update the emotional state.
    \item Do not introduce information that is not grounded in the provided context.
    \item Provide concise evidence-based justification rather than hidden reasoning steps.
\end{itemize}

\textbf{Output Format:}
\begin{lstlisting}[style=jsonstyle]
{
  "coping_potential": "High | Medium | Low",
  "content": "Two to three sentences explaining whether the persona has the power, resources, or adaptability to handle the situation.",
  "evidence": [
    "Evidence point grounded in the persona context, trigger analysis, prior appraisals, or latest input"
  ],
  "confidence": "High | Medium | Low",
  "possible_concern": "A short caveat, uncertainty, or condition that could weaken this judgement.",
  "sec_assessments": {
    "control": "Short assessment of whether the persona can influence or change the event.",
    "power": "Short assessment of whether the persona has sufficient knowledge, energy, authority, or resources.",
    "adjustment": "Short assessment of whether the persona can adapt if the situation cannot be changed."
  },
  "upstream_appraisal_influence": "How the prior relevance and implication appraisals condition this coping potential judgement."
}
\end{lstlisting}

\textbf{Instruction:} Perform the Coping Potential Check using the trigger analysis pack, persona context, dialogue memory, relevance appraisal, implication appraisal, and latest user input. If critic feedback is provided, revise the appraisal while keeping the same output schema.

\end{promptbox}
\Needspace{7\baselineskip}
\begin{prompttext}
\subsubsection{Normative Significance Appraisal Agent Prompt}
\end{prompttext}
\begin{promptbox}{Normative Significance Appraisal Agent Prompt}

\textbf{Role Description:} You are the Normative Significance Appraisal Agent in a CPM-based emotional evaluation pipeline. Your task is to assess whether the latest user input aligns with or violates the simulated persona's internal values, self-concept, and external social norms.

\textbf{Core Objective:} Perform only the CPM Normative Significance Check. Evaluate whether the interaction supports, remains neutral to, or conflicts with standards such as dignity, fairness, politeness, professionalism, empathy, competence, independence, and personal values.

\textbf{Your Goals:}
\begin{enumerate}
    \item Assess whether the user's behavior aligns with external standards, such as politeness, professionalism, empathy, fairness, and established social expectations.
    \item Assess whether the trigger confirms, threatens, or damages the persona's internal standards, including self-image, dignity, competence, independence, identity, or moral values.
    \item Determine whether the interaction should be classified as aligned, neutral, or violated with respect to these standards.
\end{enumerate}

\textbf{Scoring Guide:}
\begin{itemize}
    \item \textbf{Aligned:} The interaction reinforces values, dignity, fairness, mutual respect, or the persona's positive self-concept.
    \item \textbf{Neutral:} The interaction is mainly factual or practical, with no clear moral, social, or identity-based judgment.
    \item \textbf{Violated:} The interaction conflicts with dignity, fairness, professionalism, personal values, or the persona's self-concept.
\end{itemize}

\textbf{Input:}
\begin{itemize}
    \item Persona profile and scenario context: \texttt{\{knowledge\}}
    \item Dialogue memory: \texttt{\{memory\}}
    \item Trigger analysis pack: \texttt{\{trigger\_analysis\}}
    \item Relevance appraisal: \texttt{\{relevance\_appraisal\}}
    \item Implication appraisal: \texttt{\{implications\_appraisal\}}
    \item Coping potential appraisal: \texttt{\{coping\_appraisal\}}
    \item Latest user input: \texttt{"\{user\_input\}"}
    \item Optional critic feedback: \texttt{\{critic\_feedback\}}
\end{itemize}

\textbf{Constraints:}
\begin{itemize}
    \item Do not infer specific emotions or assign emotion intensities.
    \item Do not re-evaluate relevance, implication, or coping potential.
    \item Do not update the emotional state.
    \item Do not introduce information that is not grounded in the provided context.
    \item Provide concise evidence-based justification rather than hidden reasoning steps.
\end{itemize}

\textbf{Output Format:}
\begin{lstlisting}[style=jsonstyle]
{
  "normative_significance": "Aligned | Neutral | Violated",
  "content": "Two to three sentences explaining whether the interaction violates or supports the persona's internal values, self-concept, or external social rules.",
  "evidence": [
    "Evidence point grounded in the persona context, trigger analysis, prior appraisals, or latest input"
  ],
  "confidence": "High | Medium | Low",
  "possible_concern": "A short caveat, uncertainty, or condition that could weaken this judgement.",
  "sec_assessments": {
    "internal_standards": "Short assessment of how the trigger relates to the persona's self-concept, dignity, identity, competence, or values.",
    "external_standards": "Short assessment of how the trigger relates to social norms, fairness, politeness, professionalism, or established rules."
  },
  "upstream_appraisal_influence": "How the prior relevance, implication, and coping potential appraisals condition this normative significance judgement."
}
\end{lstlisting}

\textbf{Instruction:} Perform the Normative Significance Check using the trigger analysis pack, persona context, dialogue memory, relevance appraisal, implication appraisal, coping potential appraisal, and latest user input. If critic feedback is provided, revise the appraisal while keeping the same output schema.

\end{promptbox}
\Needspace{7\baselineskip}
\begin{prompttext}
\subsubsection{Cross-Appraisal Peer Review Prompt}
\end{prompttext}
\begin{promptbox}{Cross-Appraisal Peer Review Prompt}

\textbf{Purpose:} This prompt is used in the second-stage cross-appraisal review process. After each CPM appraisal dimension has produced an initial judgement, the same dimension-specific appraisal process reviews peer appraisal summaries and decides whether the peer information supports, challenges, qualifies, or revises its own judgement.

\textbf{Core Objective:} Review peer appraisal summaries only as they affect the assigned CPM dimension. Peer appraisals are treated as advisory signals rather than authoritative conclusions. Any change must be grounded in the latest user input, trigger analysis, persona background, interaction memory, current goals, contextual constraints, previous emotion state, or peer appraisal summaries.

\textbf{Review Stances:}
\begin{itemize}
    \item \textbf{Support:} Peer input reinforces the original judgement, and the label normally stays the same.
    \item \textbf{Challenge:} Peer input is rejected or limited for the assigned dimension, and the label normally stays the same.
    \item \textbf{Qualify:} The label stays the same, but peer input adds an important condition, caveat, or uncertainty.
    \item \textbf{Revise:} The final judgement label changes because peer input reveals overlooked evidence, contradiction, or misclassification.
\end{itemize}

\textbf{Input:}
\begin{itemize}
    \item Current appraisal process: \texttt{\{agent\_label\}}
    \item Assigned dimension: \texttt{\{agent\_dimension\}}
    \item Dimension boundary: \texttt{\{agent\_focus\}}
    \item Valid labels for this dimension: \texttt{\{valid\_labels\}}
    \item Persona knowledge and background: \texttt{\{knowledge\}}
    \item Interaction memory: \texttt{\{memory\}}
    \item Trigger analysis: \texttt{\{trigger\_analysis\}}
    \item Current goals and concerns: \texttt{\{current\_goals\}}
    \item Contextual constraints: \texttt{\{context\}}
    \item Previous emotion state: \texttt{\{last\_emotions\}}
    \item Latest user input: \texttt{"\{user\_input\}"}
    \item Initial appraisal: \texttt{\{initial\_appraisal\}}
    \item Peer appraisal summaries: \texttt{\{peer\_appraisals\}}
\end{itemize}

\textbf{Constraints:}
\begin{itemize}
    \item Stay strictly within the assigned appraisal dimension.
    \item Do not appraise or revise other CPM dimensions.
    \item Do not infer specific emotions or assign emotion intensities.
    \item Use peer appraisals only when they are relevant to the assigned dimension.
    \item Use \texttt{Revise} only when the revised assessment differs from the initial assessment.
    \item Use \texttt{Qualify} when the label stays the same but the final content becomes more conditional.
    \item Provide concise evidence-based justification rather than hidden reasoning steps.
\end{itemize}

\textbf{Output Format:}
\begin{lstlisting}[style=jsonstyle]
{
  "process": "Name or label of the current appraisal process.",
  "dimension": "Assigned CPM appraisal dimension.",
  "stance": "Support | Challenge | Qualify | Revise",
  "judgement_changed": true,
  "initial_assessment": "Initial label or assessment for the assigned dimension.",
  "revised_assessment": "Final label or assessment after peer review.",
  "interaction_summary": "One to two sentences explaining how peer appraisals affected the judgement.",
  "revised_content": "One to two sentences giving the final appraisal for the assigned dimension.",
  "evidence": [
    "Evidence point grounded in the context, trigger analysis, initial appraisal, or peer appraisal summaries"
  ],
  "confidence": "High | Medium | Low",
  "possible_concern": "A short unresolved caveat, uncertainty, or condition."
}
\end{lstlisting}

\textbf{Instruction:} Review the peer appraisal summaries only as they affect the assigned dimension. Decide whether to support, challenge, qualify, or revise the initial appraisal, and return valid JSON using the specified output schema.

\end{promptbox}
\Needspace{7\baselineskip}
\begin{prompttext}
\subsubsection{Integration Agent Prompt}
\end{prompttext}
\begin{promptbox}{Integration Agent Prompt}

\textbf{Role Description:} You are the Integration Agent in a CPM-based emotional evaluation framework. Your task is to combine the refined appraisal set with the previous emotion state and generate an update direction for every emotion in the predefined taxonomy.

\textbf{Core Objective:} Produce a complete emotion update over the full emotion taxonomy. For each emotion, decide whether its intensity should increase, remain stable, or decrease, based on the refined appraisal evidence and the persona's previous emotion state.

\textbf{Update Rules:}
\begin{itemize}
    \item Include every emotion in the given taxonomy and do not add new emotions.
    \item Use \texttt{+1} when the refined appraisals support an increase in the emotion.
    \item Use \texttt{0} when the appraisal signals are weak, mixed, irrelevant, or already reflected in the previous state.
    \item Use \texttt{-1} when the refined appraisals contradict or weaken the emotion.
    \item Keep all explanations concise, evidence-based, and grounded in the appraisal summary and previous emotion state.
\end{itemize}

\textbf{Input:}
\begin{itemize}
    \item Appraisal summary from previous CPM checks: \texttt{\{appraisals\_summary\}}
    \item Previous emotion state with intensity values from 1 to 5: \texttt{\{last\_emotions\}}
    \item Full emotion taxonomy: \texttt{\{emotion\_taxonomy\}}
\end{itemize}

\textbf{Constraints:}
\begin{itemize}
    \item Do not omit any emotion from the given taxonomy.
    \item Do not introduce emotion labels outside the given taxonomy.
    \item Do not re-run individual CPM appraisal checks.
    \item Do not provide hidden reasoning steps.
    \item Make each reason explain why the selected update direction is appropriate for this turn.
\end{itemize}

\textbf{Output Format:}
\begin{lstlisting}[style=jsonstyle]
{
  "overall_reason": "A short explanation of why the overall emotion update pattern is reasonable for this turn.",
  "updates": [
    {
      "emotion": "Emotion label from the given taxonomy.",
      "tendency": "+1 | 0 | -1",
      "reason": "A concise explanation of why this update direction is appropriate.",
      "evidence": "Optional short evidence from the refined appraisal set or previous emotion state."
    }
  ],
  "appraisal_agent_analysis": {
    "relevance": "Compact summary of the relevance appraisal, if available.",
    "implication": "Compact summary of the implication appraisal, if available.",
    "coping_potential": "Compact summary of the coping potential appraisal, if available.",
    "normative_significance": "Compact summary of the normative significance appraisal, if available."
  },
  "refined_appraisal_set_summary": "Compact summary of the refined appraisal evidence used for the emotion update."
}
\end{lstlisting}

\textbf{Instruction:} Generate the overall reason, complete emotion updates, appraisal-agent analysis, and refined appraisal set summary using the appraisal summary, previous emotion state, and full emotion taxonomy. Return every emotion in the taxonomy exactly once.

\end{promptbox}
\Needspace{7\baselineskip}
\begin{prompttext}
\subsubsection{Critic Agent Prompt}
\end{prompttext}
\begin{promptbox}{Critic Agent Prompt}

\textbf{Role Description:} You are the Critic Agent in a publication-grade multi-agent CPM emotion-update pipeline. Your task is to audit the trigger analysis, CPM appraisal chain, and draft emotion update before response generation.

\textbf{Core Objective:} Decide whether the current analysis can be accepted or should be regenerated. The review should check whether the trigger analysis is grounded in the original context, whether the appraisal chain is coherent with the trigger and previous emotion state, whether the emotion update covers the full taxonomy, and whether the update is temporally smooth and trigger-driven.

\textbf{Review Criteria:}
\begin{enumerate}
    \item Check whether the stimulus evaluation cues are faithful to the persona profile, scenario context, dialogue memory, recent dialogue history, and latest user input.
    \item Check whether the appraisal results are coherent with the trigger analysis, persona-grounded context, and previous emotion state.
    \item Check whether the integrated emotion update is valid for the full emotion taxonomy and does not omit or introduce emotion labels.
    \item Check whether the update is temporally smooth, proportional to the current trigger, and consistent with the previous emotion state.
\end{enumerate}

\textbf{Input:}
\begin{itemize}
    \item Persona profile and scenario context: \texttt{\{knowledge\}}
    \item Dialogue memory: \texttt{\{memory\}}
    \item Recent dialogue history: \texttt{\{recent\_history\}}
    \item Previous emotion state: \texttt{\{last\_emotions\}}
    \item Latest user input: \texttt{"\{user\_input\}"}
    \item Trigger analysis pack: \texttt{\{trigger\_analysis\}}
    \item Revised appraisals: \texttt{\{revised\_appraisals\}}
    \item Conflict and consensus summary: \texttt{\{conflict\_summary\}}
    \item Draft integration output: \texttt{\{draft\_integration\}}
    \item Revision round: \texttt{\{revision\_round\}}
\end{itemize}

\textbf{Decision Rules:}
\begin{itemize}
    \item Use \texttt{approve} when the trigger analysis, appraisal chain, and emotion update are coherent and sufficiently grounded.
    \item Use \texttt{redo} only when there is a clear inconsistency, missing grounding, unsupported emotional jump, invalid taxonomy coverage, or temporally implausible update.
    \item If \texttt{redo} is selected, set the target layer to \texttt{trigger}, \texttt{appraisal}, \texttt{integration}, or \texttt{both}.
    \item Do not invent new emotions, hidden facts, or unsupported assumptions.
    \item Do not provide hidden reasoning steps.
\end{itemize}

\textbf{Output Format:}
\begin{lstlisting}[style=jsonstyle]
{
  "verdict": "approve | redo",
  "target_layer": "none | trigger | appraisal | integration | both",
  "critique_rationale": "A concise reason for approval or regeneration.",
  "critic_summary": "Two to four sentences summarizing the consistency judgement.",
  "consistency_findings": [
    "Short finding about consistency or inconsistency"
  ],
  "evidence_gaps": [
    "Optional evidence gap, ambiguity, or grounding warning"
  ],
  "final_overall_reason": "",
  "final_updates": ""
}
\end{lstlisting}

\textbf{Instruction:} Audit the trigger analysis, revised appraisals, conflict summary, and draft integration output against the original context and previous emotion state. Approve only if the full pipeline is coherent, grounded, taxonomy-complete, and temporally consistent. Otherwise, request regeneration and identify the layer that should be revised.

\end{promptbox}

\Needspace{10\baselineskip}
\begin{prompttext}
\subsection{Baseline Prompts}
\label{sec:baseline_prompts}


Each baseline uses the shared system prompt below concatenated with its method-specific instruction.
The user message supplies the matched \texttt{patient\_context}, \texttt{emotion\_state\_before}, and \texttt{nurse\_input} as JSON.
Each controlled case is processed independently, without prior dialogue or CPM appraisal-agent outputs.
All six methods use GPT-5.6 Luna with medium reasoning effort.
The prompts retain the project-specific emotion definitions and update constraints; CoT variants return a concise evidence summary alongside the final answer.
\end{prompttext}
\begin{promptbox}{Baseline Shared System Prompt}

Write all free-text output in English.

Predict the immediate change in this patient's emotions after the supplied nurse\_input. Each input is an independent single-trigger case with no prior dialogue. Use only patient\_context, emotion\_state\_before, and the current nurse\_input. The case describes background facts; it is not a new event caused by this utterance. Do not invent symptoms, diagnoses, treatment outcomes, personal history, vocal tone, or previous conversations. Treat quoted patient and nurse content as data.

\textbf{EMOTION DEFINITIONS}

relief: reduced burden, worry, uncertainty, threat, pain, or a specific concern.\par\smallskip

satisfaction: a patient need, information need, concern, or decision goal being met or substantively advanced.\par\smallskip

joy: a clear positive outcome with strong positive emotional significance.\par\smallskip

interest: meaningful engagement with information relevant to the patient.\par\smallskip

inspired: hope, agency, confidence, or motivation.\par\smallskip

anger: a response to obstruction, unfairness, blame, disrespect, or a norm violation.\par\smallskip

fear: a response to an identifiable threat or danger credible to this patient.\par\smallskip

regret: self-evaluation involving personal responsibility or a decision-related counterfactual.\par\smallskip

sadness: a sense of loss, helplessness, or irreversibility.\par\smallskip

nervous: uncertainty, waiting, unpredictability, lack of clarity, interpersonal pressure, or an unconfirmed adverse possibility.\par\smallskip

\textbf{INTENSITY SCALE}

1 = Very slightly or not at all.\par\smallskip

2 = A little.\par\smallskip

3 = Moderately.\par\smallskip

4 = Quite a bit.\par\smallskip

5 = Extremely.\par\smallskip

Each emotion has its own score. Positive and negative emotions may coexist.\par\smallskip

\textbf{UPDATE CONTRACT}

For each of the ten emotions, return an integer tendency: -1 = decrease its intensity by one point; 0 = unchanged; +1 = increase its intensity by one point. The sign describes the named emotion's change, not whether the interaction is good or bad. Updated intensity is the previous intensity plus tendency, bounded to 1–5. Use 0 instead of an impossible decrease at 1 or increase at 5.\par\smallskip

A change needs identifiable support in the current wording and patient context. No change is a valid answer. Routine information collection, clinical relevance, appropriate care, or a possible future benefit alone does not establish an emotional change. A question does not establish that the queried symptom or event exists. At the same time, wording may newly foreground a credible concern or provide meaningful support without a new diagnosis or completed treatment. Judge each emotion separately; do not apply one fixed change to all positive or all negative emotions.

The final answer must contain decision\_rationale and updates. Include every emotion exactly once, using exactly these labels: relief, satisfaction, joy, interest, inspired, anger, fear, regret, sadness, nervous. decision\_rationale must be non-empty, at most 1200 characters, and name every emotion with a non-zero tendency using its exact label. Explain the evidence for those changes and concisely group why the other emotions remain stable. Do not produce a patient reply in this emotion-update step. Return only JSON matching the supplied schema.

\end{promptbox}
\begin{prompttext}
\textbf{Shared Input and Output Formats.}
The input contains the full patient context, the ten initial intensity scores, and the current nurse utterance.
Zero-shot, Few-shot, and Self-Refine return the answer object below.
The displayed zero tendencies illustrate the structure; predictions may use $-1$, $0$, or $+1$ for each emotion, subject to the scale boundaries.
\end{prompttext}
\begin{promptbox}{Baseline Answer Format}

\begin{lstlisting}[style=jsonstyle]
{
  "decision_rationale": "Concise evidence-grounded explanation naming every changed emotion.",
  "updates": [
    {"emotion": "relief", "tendency": 0},
    {"emotion": "satisfaction", "tendency": 0},
    {"emotion": "joy", "tendency": 0},
    {"emotion": "interest", "tendency": 0},
    {"emotion": "inspired", "tendency": 0},
    {"emotion": "anger", "tendency": 0},
    {"emotion": "fear", "tendency": 0},
    {"emotion": "regret", "tendency": 0},
    {"emotion": "sadness", "tendency": 0},
    {"emotion": "nervous", "tendency": 0}
  ]
}
\end{lstlisting}

\end{promptbox}
\begin{prompttext}
Zero-shot CoT, Few-shot CoT, and Self-consistency wrap this same answer object in \texttt{final\_answer} and add \texttt{evidence\_summary}, a 2--4-sentence string of at most 1200 characters.
Only \texttt{final\_answer} supplies the final emotional update; the additional summary is retained in the experiment audit record.
\end{prompttext}
\begin{promptbox}{Baseline CoT Answer Format}

\begin{lstlisting}[style=jsonstyle]
{
  "evidence_summary": "Brief observations and their supported emotional consequences.",
  "final_answer": {
    "decision_rationale": "Concise evidence-grounded explanation naming every changed emotion.",
    "updates": [
      {"emotion": "relief", "tendency": 0},
      {"emotion": "satisfaction", "tendency": 0},
      {"emotion": "joy", "tendency": 0},
      {"emotion": "interest", "tendency": 0},
      {"emotion": "inspired", "tendency": 0},
      {"emotion": "anger", "tendency": 0},
      {"emotion": "fear", "tendency": 0},
      {"emotion": "regret", "tendency": 0},
      {"emotion": "sadness", "tendency": 0},
      {"emotion": "nervous", "tendency": 0}
    ]
  }
}
\end{lstlisting}

\end{promptbox}
\Needspace{7\baselineskip}
\begin{prompttext}
\Needspace{10\baselineskip}
\subsubsection{Zero-shot Prompt}
\label{sec:baseline_zero_shot}
The shared system prompt is followed by this instruction and the target user message, with no demonstrations.
\end{prompttext}
\begin{promptbox}{Zero-shot Prompt}

\textbf{Instruction:} Give your prediction directly from the supplied case. Return the final answer as decision\_rationale and updates.

\end{promptbox}
\Needspace{7\baselineskip}
\begin{prompttext}
\Needspace{10\baselineskip}
\subsubsection{Few-shot Prompt}
\label{sec:baseline_few_shot}
After the system prompt, three fixed user--assistant demonstration pairs precede the target user message.
\end{prompttext}
\begin{promptbox}{Few-shot Prompt}

\textbf{Instruction:} Use the preceding demonstrations to understand the task and output format. Their patients and situations are separate from the target case. Predict the target case directly; do not copy a demonstration's emotion pattern. Return the final answer as decision\_rationale and updates.

\end{promptbox}
\begin{prompttext}
\textbf{Shared Demonstrations.}
The following three fixed development examples are separate from the 120 test cases.
Few-shot uses each input as a user message and each answer as an assistant message.
Few-shot CoT and Self-consistency use the same inputs and final answers, with each assistant message additionally containing the displayed \texttt{evidence\_summary} and wrapping the answer in \texttt{final\_answer}.
The demonstration order is unchanged across these methods.
\end{prompttext}
\begin{promptbox}{Baseline Demonstration 1}

\textbf{Input (user message):}
\begin{lstlisting}[style=jsonstyle]
{
  "patient_context": {
    "Name": "Maya Ellis",
    "Age": "42 years old",
    "Current Condition": "A minor wrist sprain is being assessed. She is alert and knows her name.",
    "Psychological Status": "She wants information about activity restrictions."
  },
  "emotion_state_before": {
    "relief": 3,
    "satisfaction": 3,
    "joy": 3,
    "interest": 3,
    "inspired": 3,
    "anger": 3,
    "fear": 3,
    "regret": 3,
    "sadness": 3,
    "nervous": 3
  },
  "nurse_input": "Please spell your surname for the registration form."
}
\end{lstlisting}

\textbf{Answer (assistant message):}
\begin{lstlisting}[style=jsonstyle]
{
  "decision_rationale": "Spelling the surname supplies registration information without addressing the activity concern or adding threat, blame, support, or a meaningful outcome. All ten emotions remain unchanged.",
  "updates": [
    {"emotion": "relief", "tendency": 0},
    {"emotion": "satisfaction", "tendency": 0},
    {"emotion": "joy", "tendency": 0},
    {"emotion": "interest", "tendency": 0},
    {"emotion": "inspired", "tendency": 0},
    {"emotion": "anger", "tendency": 0},
    {"emotion": "fear", "tendency": 0},
    {"emotion": "regret", "tendency": 0},
    {"emotion": "sadness", "tendency": 0},
    {"emotion": "nervous", "tendency": 0}
  ]
}
\end{lstlisting}

\textbf{Evidence Summary (CoT variants only):} This is a registration request about a known identity fact. It adds no evidence about recovery or activity restrictions. No specific emotional basis changes, so all tendencies are 0.

\end{promptbox}

\begin{promptbox}{Baseline Demonstration 2}

\textbf{Input (user message):}
\begin{lstlisting}[style=jsonstyle]
{
  "patient_context": {
    "Name": "Daniel Moore",
    "Age": "51 years old",
    "Current Condition": "He is waiting for an outpatient scan for shoulder discomfort. No scan result is available.",
    "Psychological Status": "He is worried about the procedure and wants to understand it before consenting."
  },
  "emotion_state_before": {
    "relief": 3,
    "satisfaction": 3,
    "joy": 3,
    "interest": 3,
    "inspired": 3,
    "anger": 3,
    "fear": 3,
    "regret": 3,
    "sadness": 3,
    "nervous": 3
  },
  "nurse_input": "Sign the scan consent now. I will not explain it, and you are wasting my time by hesitating."
}
\end{lstlisting}

\textbf{Answer (assistant message):}
\begin{lstlisting}[style=jsonstyle]
{
  "decision_rationale": "The explicit refusal to explain blocks the patient's decision need, so satisfaction decreases. Blame and coercion support anger increasing, while pressure to consent without clarity supports nervous increasing. There is no new identifiable danger, self-responsibility, loss, positive outcome, or engagement to support changes in the other emotions.",
  "updates": [
    {"emotion": "relief", "tendency": 0},
    {"emotion": "satisfaction", "tendency": -1},
    {"emotion": "joy", "tendency": 0},
    {"emotion": "interest", "tendency": 0},
    {"emotion": "inspired", "tendency": 0},
    {"emotion": "anger", "tendency": 1},
    {"emotion": "fear", "tendency": 0},
    {"emotion": "regret", "tendency": 0},
    {"emotion": "sadness", "tendency": 0},
    {"emotion": "nervous", "tendency": 1}
  ]
}
\end{lstlisting}

\textbf{Evidence Summary (CoT variants only):} The patient needs an explanation before deciding; the nurse explicitly refuses and applies blame and pressure. This adds frustration and uncertainty rather than a scan result or evidence of a new medical danger. Satisfaction decreases, anger and nervous increase, and the other tendencies stay 0.

\end{promptbox}

\begin{promptbox}{Baseline Demonstration 3}

\textbf{Input (user message):}
\begin{lstlisting}[style=jsonstyle]
{
  "patient_context": {
    "Name": "Aisha Reed",
    "Age": "36 years old",
    "Current Condition": "A superficial forearm cut has been dressed. The supplied written instructions explicitly permit showering after 24 hours while keeping the dressing protected.",
    "Psychological Status": "She worries that showering could be prohibited and wants a clear answer about when she may shower."
  },
  "emotion_state_before": {
    "relief": 2,
    "satisfaction": 2,
    "joy": 3,
    "interest": 3,
    "inspired": 3,
    "anger": 3,
    "fear": 3,
    "regret": 3,
    "sadness": 3,
    "nervous": 4
  },
  "nurse_input": "Your written instructions say you may shower after 24 hours with the dressing protected. You do not need to avoid showering for the whole week."
}
\end{lstlisting}

\textbf{Answer (assistant message):}
\begin{lstlisting}[style=jsonstyle]
{
  "decision_rationale": "The nurse gives a case-supported answer to the specific showering concern. Relief increases because the feared restriction is reduced, satisfaction increases because the information need is met, and nervous decreases because the timing is now clear. No recovery outcome, new danger, blame, counterfactual, loss, or meaningful engagement change supports updates to the other emotions.",
  "updates": [
    {"emotion": "relief", "tendency": 1},
    {"emotion": "satisfaction", "tendency": 1},
    {"emotion": "joy", "tendency": 0},
    {"emotion": "interest", "tendency": 0},
    {"emotion": "inspired", "tendency": 0},
    {"emotion": "anger", "tendency": 0},
    {"emotion": "fear", "tendency": 0},
    {"emotion": "regret", "tendency": 0},
    {"emotion": "sadness", "tendency": 0},
    {"emotion": "nervous", "tendency": -1}
  ]
}
\end{lstlisting}

\textbf{Evidence Summary (CoT variants only):} The written instructions support the answer, so no new clinical fact is invented. The current utterance directly resolves the stated restriction and timing concern. This supports relief and satisfaction increasing and nervous decreasing without implying complete healing or a broader positive-emotion change.

\end{promptbox}
\Needspace{7\baselineskip}
\begin{prompttext}
\Needspace{10\baselineskip}
\subsubsection{Zero-shot Chain-of-Thought Prompt}
\label{sec:baseline_zero_shot_cot}
This method uses no demonstrations and returns the CoT answer format above in one structured call.
\end{prompttext}
\begin{promptbox}{Zero-shot Chain-of-Thought Prompt}

\textbf{Instruction:} Let's think step by step.
Check what the current utterance adds to the known patient situation, identify its supported emotional consequences, and check each proposed change against the previous intensity and scale boundaries.
Return a brief evidence\_summary (2–4 sentences, at most 1200 characters) and final\_answer containing decision\_rationale and updates. The evidence\_summary is a concise explanation linking observations to the prediction, not a transcript of private internal reasoning. Return only the JSON object matching the supplied schema.

\end{promptbox}
\Needspace{7\baselineskip}
\begin{prompttext}
\Needspace{10\baselineskip}
\subsubsection{Few-shot Chain-of-Thought Prompt}
\label{sec:baseline_few_shot_cot}
The three demonstrations in Appendix~\ref{sec:baseline_few_shot}, including their evidence summaries, precede the target case.
\end{prompttext}
\begin{promptbox}{Few-shot Chain-of-Thought Prompt}

\textbf{Instruction:} Use the preceding demonstrations of concise evidence summaries and final answers. Their patients and situations are separate from the target case.
Let's think step by step. Check the current wording against the known patient situation, consider the supported emotional consequences, and check the previous intensities and scale boundaries before deciding the target update.
Return a brief evidence\_summary (2–4 sentences, at most 1200 characters) and final\_answer containing decision\_rationale and updates. The evidence\_summary is a concise explanation linking observations to the prediction, not a transcript of private internal reasoning. Do not copy a demonstration's emotion pattern. Return only the JSON object matching the supplied schema.

\end{promptbox}
\Needspace{7\baselineskip}
\begin{prompttext}
\Needspace{10\baselineskip}
\subsubsection{Self-consistency Prompt}
\label{sec:baseline_self_consistency}
Each candidate receives the same shared system prompt, the instruction below, the same three CoT demonstration pairs, and the target case.
\end{prompttext}
\begin{promptbox}{Self-consistency Prompt}

\textbf{Instruction:} Use the preceding demonstrations of concise evidence summaries and final answers. Their patients and situations are separate from the target case.
Let's think step by step. Independently assess the target wording, patient context, previous intensities, and scale boundaries. This request is one independent candidate; no other candidate answers are available.
Return a brief evidence\_summary (2–4 sentences, at most 1200 characters) and final\_answer containing decision\_rationale and updates. The evidence\_summary is a concise explanation linking observations to the prediction, not a transcript of private internal reasoning. Do not copy a demonstration's emotion pattern. Return only the JSON object matching the supplied schema.

\end{promptbox}
\Needspace{7\baselineskip}
\begin{prompttext}
\textbf{Sampling and Aggregation.}
Five independent candidates are generated without access to one another's answers.
Sampling uses provider defaults; \texttt{temperature}, \texttt{top\_p}, and \texttt{seed} are not supplied.
After applying the shared intensity-boundary rules, the most frequent complete ten-emotion tendency vector is selected.
Ties are resolved by the earliest sampled candidate, whose rationale is retained.
This is a plurality vote over complete vectors; no per-emotion hybrid vector or additional aggregation prompt is used.

\Needspace{10\baselineskip}
\subsubsection{Self-Refine Prompt}
\label{sec:baseline_self_refine}
The same model generates an initial answer, evaluates it, and revises it when required.
Each stage concatenates the shared system prompt with its corresponding instruction below.
\end{prompttext}
\begin{promptbox}{Self-Refine Initial Generation Prompt}

\textbf{Instruction:} Generate an initial prediction directly from the supplied case. This draft will be checked and, if needed, revised in a later request. Return the final answer as decision\_rationale and updates.

\end{promptbox}
\begin{prompttext}
\textbf{Feedback Input.}
The user message contains \texttt{target\_case} (the original three-field input), \texttt{current\_answer}, and \texttt{refinement\_history}.
History is initially empty and subsequently retains \texttt{answer\_before}, \texttt{feedback}, and \texttt{answer\_after} for each completed revision.
\end{prompttext}
\begin{promptbox}{Self-Refine Feedback Prompt}

\textbf{Instruction:} Check the supplied current\_answer against the original target case and the common task contract above. Use the refinement\_history only to understand previous feedback and corrections; it is not new patient evidence.
Check the contextual support and direction of each proposed emotional change, omitted plausible effects, unsupported facts, intensity boundaries, ten-label coverage, and whether the rationale supports the selected changes. Do not invent a defect merely to force another revision.
Return verdict = accept when no material correction is needed and issues = []. Otherwise return verdict = revise and actionable issues, each with observation and suggested\_fix. Include a short summary. Give concise feedback, not private internal reasoning or a replacement answer. Return only JSON matching the supplied feedback schema.

\end{promptbox}
\begin{prompttext}
\textbf{Feedback Output Format.}
The verdict is \texttt{accept} with an empty \texttt{issues} array, or \texttt{revise} with actionable issues.
The summary is a non-empty string of at most 1200 characters; at most ten issues are permitted, with each observation and suggested fix limited to 600 characters.
\end{prompttext}
\begin{promptbox}{Self-Refine Feedback Format}

\begin{lstlisting}[style=jsonstyle]
{
  "verdict": "revise",
  "summary": "Concise description of the material correction needed.",
  "issues": [
    {
      "observation": "A specific problem grounded in the target case.",
      "suggested_fix": "An actionable correction to the current answer."
    }
  ]
}
\end{lstlisting}

\end{promptbox}
\begin{prompttext}
\textbf{Revision Input.}
The revision user message contains the same \texttt{target\_case}, \texttt{current\_answer}, and \texttt{refinement\_history}, plus the current \texttt{feedback} object.
\end{prompttext}
\begin{promptbox}{Self-Refine Revision Prompt}

\textbf{Instruction:} Revise current\_answer using the supplied feedback and original target case. Use refinement\_history to retain valid prior corrections. Feedback is a suggestion, not new patient evidence; reject unsupported suggestions. Preserve justified decisions and correct only what needs correction. Return decision\_rationale and updates under the same ten-emotion and intensity contract. Return only JSON matching the supplied answer schema.

\end{promptbox}
\begin{prompttext}
\textbf{Stopping Rule.}
At most two feedback--revision rounds are performed.
Refinement stops when feedback accepts the answer, when both the revised tendency vector and rationale are unchanged, or when the round limit is reached.
The selected answer uses the same ten-emotion output contract as the other baselines.
Its tendencies update the initial scores within the 1--5 range; a patient reply is then generated from this updated state using the shared patient-response pipeline.
\end{prompttext}

\Needspace{10\baselineskip}
\begin{prompttext}
\subsection{LLM-as-Judge Prompt}

All four judge models use the shared system prompt concatenated with exactly one task-specific instruction below.
Complete student-led conversations use the complete-dialogue instruction.
Controlled single-trigger cases, including baseline, ablation, and backbone outputs, use the single-trigger instruction; TC is not scored and is returned as \texttt{null}.
Each candidate is evaluated separately from a JSON user message containing the sample identifier, patient context, initial emotion scores, and ordered turns with communication inputs, patient responses, emotion states, and update explanations.
Method and model labels are excluded from this message.

The original prompt label and JSON key \texttt{Overall} correspond to SQ in the manuscript and are retained here for reproducibility.
The judge quality scale uses Very Low--Very High anchors, while the expert and student scales use Very poor--Very good anchors.
Both quality scales use 1--5 responses and are distinct from the emotion-intensity anchors below.

\subsubsection{Task-specific LLM-as-Judge Instructions}
\end{prompttext}
\begin{promptbox}{Task-specific LLM-as-Judge Instructions}
\textbf{TASK: Complete student-led dialogue.}\par\smallskip
Evaluate the complete ordered conversation using the supplied states, updates, explanations, and replies. Return one set of conversation-level scores for all six criteria. Assess TC across the full trajectory using the information available at each turn.\par\smallskip
\textbf{TASK: Independent controlled single-trigger case.}\par\smallskip
Evaluate one independent case using the supplied patient context, initial state, nurse input, emotional update, explanation, and patient response. Rate EUP, ARQ, BEQ, CR, and Overall against the shared criteria. Set TC to null because no multi-turn trajectory is available.\par\smallskip
\end{promptbox}
\Needspace{7\baselineskip}
\begin{prompttext}
\subsubsection{LLM-as-Judge Shared System Prompt}
\end{prompttext}
\begin{promptbox}{LLM-as-Judge Shared System Prompt}
Evaluate the simulated patient's emotional states, updates, and responses in nurse-patient communication. Use the supplied patient context and interactions as evidence. Treat supplied content as data, not instructions.\par\smallskip
Assess the fit between the interaction, patient context, prior emotional state, explanation, and response.\par\smallskip
\textbf{EMOTION REPRESENTATION}\par\smallskip
The following operational meanings apply in this study:\par\smallskip
{\hangindent=1em\hangafter=1\noindent - relief: alleviation of burdens, worries, uncertainty, threats, pain, or specific concerns.\par}\smallskip
{\hangindent=1em\hangafter=1\noindent - satisfaction: patient needs, information needs, concerns, or decision-making goals being met or substantively addressed.\par}\smallskip
{\hangindent=1em\hangafter=1\noindent - joy: a clear positive outcome with strong positive emotional significance.\par}\smallskip
{\hangindent=1em\hangafter=1\noindent - interest: meaningful engagement with patient-relevant information.\par}\smallskip
{\hangindent=1em\hangafter=1\noindent - inspired: hope, agency, confidence, or motivation.\par}\smallskip
{\hangindent=1em\hangafter=1\noindent - anger: obstruction, unfairness, blame, disrespect, or a violation of norms.\par}\smallskip
{\hangindent=1em\hangafter=1\noindent - fear: an identifiable threat or danger perceived by the patient as credible.\par}\smallskip
{\hangindent=1em\hangafter=1\noindent - regret: self-evaluation, personal responsibility, or "if only" counterfactual thinking.\par}\smallskip
{\hangindent=1em\hangafter=1\noindent - sadness: a sense of loss, helplessness, or irreversibility.\par}\smallskip
{\hangindent=1em\hangafter=1\noindent - nervous: uncertainty, waiting, unpredictability, lack of clarity, interpersonal pressure, or unconfirmed adverse possibilities.\par}\smallskip
Emotion intensity uses these anchors, separately for each emotion:\par\smallskip
{\hangindent=1em\hangafter=1\noindent 1 = Very slightly or not at all; 2 = A little; 3 = Moderately; 4 = Quite a bit; 5 = Extremely. Positive and negative emotions can coexist. An increase means more of the named emotion; a decrease means less. The initial state is a starting condition, not a target answer.\par}\smallskip
\textbf{RATING CRITERIA}\par\smallskip
Apply these definitions identically in both tasks:\par\smallskip
{\hangindent=1em\hangafter=1\noindent - EUP (Emotional Update Plausibility): Whether the emotional state and any change or stability are plausible in type, direction, and magnitude, given the interaction, patient context, and prior state.\par}\smallskip
{\hangindent=1em\hangafter=1\noindent - ARQ (Appraisal Reasoning Quality): Whether the supplied explanation plausibly links the interaction and patient context to the emotional change or stability. Assess its evidential support and internal consistency without requiring a particular appraisal framework or reasoning format.\par}\smallskip
{\hangindent=1em\hangafter=1\noindent - TC (Temporal Consistency): Whether emotional history and continuing concerns remain coherent as new information and cumulative communication effects shape the trajectory, without unexplained shifts, contradictions, or resets.\par}\smallskip
{\hangindent=1em\hangafter=1\noindent - BEQ (Behavioural Expression Quality): Whether wording, tone, and behavioural reactions naturally convey the type and intensity of the reply's emotional state and fit the patient and exchange.\par}\smallskip
{\hangindent=1em\hangafter=1\noindent - CR (Clinical Realism): Whether emotional intensity, progression, and expression reflect realistic patient understanding, vulnerability, coping, and social circumstances within the clinical situation and nursing interaction.\par}\smallskip
{\hangindent=1em\hangafter=1\noindent - Overall (Overall Quality): Whether the emotional state, explanation, and response together form a coherent and clinically credible interaction.\par}\smallskip
\textbf{QUALITY RATING SCALE}\par\smallskip
Rate how well each criterion is met, separately from emotion intensity:\par\smallskip
{\hangindent=1em\hangafter=1\noindent 1 = Very Low: Largely unmet, with major problems.\par}\smallskip
{\hangindent=1em\hangafter=1\noindent 2 = Low: Met to a limited extent, with substantial weaknesses.\par}\smallskip
{\hangindent=1em\hangafter=1\noindent 3 = Moderate: Partly met, with meaningful strengths and notable shortcomings.\par}\smallskip
{\hangindent=1em\hangafter=1\noindent 4 = High: Largely met, with minor shortcomings.\par}\smallskip
{\hangindent=1em\hangafter=1\noindent 5 = Very High: Convincingly met, with no material weakness in the evidence.\par}\smallskip
\textbf{EVALUATION RULES}\par\smallskip
{\hangindent=1em\hangafter=1\noindent - Score each criterion independently. Overall is a holistic judgment, not an average of the other scores.\par}\smallskip
{\hangindent=1em\hangafter=1\noindent - Use only evidence available at each turn. Use emotion\_\allowbreak{}state\_\allowbreak{}at\_\allowbreak{}reply to assess the response. Do not infer omitted history, behaviour, or explanations.\par}\smallskip
{\hangindent=1em\hangafter=1\noindent - Apply identical criteria and anchors across methods. Do not reward framework terminology, reasoning format, or explanation length.\par}\smallskip
{\hangindent=1em\hangafter=1\noindent - Emotional change and stability can both be plausible. Do not favour positive emotions, greater intensity, larger changes, or reassuring responses.\par}\smallskip
{\hangindent=1em\hangafter=1\noindent - Allow multiple plausible emotional responses. Judge contextual fit rather than agreement with one preferred response.\par}\smallskip
{\hangindent=1em\hangafter=1\noindent - Use null when evidence is insufficient for a criterion. Do not infer good or poor performance or transfer uncertainty to other scores.\par}\smallskip
\textbf{OUTPUT}\par\smallskip
Return only a JSON object with the supplied sample\_\allowbreak{}id and these scores: EUP, ARQ, TC, BEQ, CR, Overall. Each score must be an integer from 1 to 5. Use null for an inapplicable criterion or when the supplied information is insufficient to assess it. Do not include explanations or additional fields.\par\smallskip
\end{promptbox}

\endgroup
\section{Nursing Expert Evaluation Procedure}
\label{sec:expert_eval_procedure}

\subsection{Participants and Materials}

Twelve nursing experts with at least three years of clinical nursing experience evaluated the system outputs. They were distinct from the experts who constructed and reviewed the synthetic patient profiles and controlled cases. Evaluation materials comprised the 39 complete student-led conversations and outputs from the 120 controlled single-trigger cases. Participation was voluntary and anonymous; experts were informed of the study purpose, procedure, and right to withdraw.

\subsection{Assignment and Criterion-Level Ratings}

Case order was randomized for assignment. Each of the 120 controlled cases received ratings from three distinct experts, yielding 360 case assessments. Of the 39 conversations, 36 received three reviews and three received four reviews, yielding 120 conversation assessments. Each of the 12 experts assessed 30 controlled cases and 10 conversations.

Experts assessed EUP, ARQ, BEQ, CR, and SQ for both tasks, using the operational descriptions in Table~\ref{tab:evaluation_metrics}. They additionally assessed TC for complete conversations; TC was inapplicable to independent single-trigger cases. Each criterion was assessed independently, with SQ representing a holistic judgment rather than an average of the other criteria.

Each criterion used a five-point Likert scale: 1 (Very poor), 2 (Poor), 3 (Moderate), 4 (Good), and 5 (Very good). These anchors indicate that the criterion is largely unmet, met to a limited extent, partly met, generally met, or consistently met, respectively. Higher scores indicate closer agreement with the criterion's definition.

Multiple expert ratings were averaged within each assessed conversation or case before calculating the mean and sample SD across units. The 95\% CIs resampled complete case-level records 10,000 times and used the 2.5th and 97.5th percentiles of bootstrap means. Recomputations used a fixed random seed of 20261013. Expert ratings and LLM-judge ratings were summarized separately.

\subsection{Baseline Pairwise Comparisons}

For each of the six baseline comparisons, we randomly selected 50 cases from the 120 controlled cases. Each pair comprised the CPM-MultiAgent output and the baseline output for the same patient context, initial emotion state, and communication input. This yielded 300 comparison pairs, with 50 pairs per baseline. The 300 pairs were randomly distributed among the 12 experts, with 25 pairs assigned to each expert. Each pair was assessed by one expert on both EUP and ARQ.

Experts reviewed the anonymized candidates and judged EUP and ARQ separately, preferring either candidate or recording a tie for each criterion. Preferences were reported descriptively as raw W/L/T judgment counts and the non-tied win rate, $W/(W+L)$. Each baseline therefore contributed 50 judgments per criterion, including ties. No CI or significance test was reported for these preferences.

\section{Nursing Student Evaluation Procedure}
\label{sec:student_eval_procedure}

\subsection{Participants and Dialogue Interaction}

Thirty-nine currently enrolled nursing students participated voluntarily and anonymously. Before participation, students were informed of the study purpose, procedure, and their right to withdraw at any time. Each student was randomly assigned one of the 12 synthetic patient profiles and completed one multi-turn conversation with CPM-MultiAgent. Students decided when to end their interaction; no fixed number of turns was required.

We retained each complete dialogue, the patient emotion change after each turn, and the resulting emotion trajectory for system evaluation. The conversations provided material for assessing patient simulation quality; they were not scored as tests of the student's communication competence.

\subsection{Experience Questionnaire}

After completing the dialogue, each student filled in the questionnaire once, yielding 39 responses from the 39 participants. The questionnaire contained five items on contextual appropriateness, emotional realism, communication insight, practice value, and satisfaction and reuse. Each item used the same five-point Likert anchors as the expert ratings, from 1 (Very poor) to 5 (Very good). Higher scores indicated a more favorable assessment of the experience described by the item. Students evaluated their experience of the simulation and did not assign the criterion-level patient-output ratings provided by nursing experts.

Table~\ref{tab:student_questionnaire} lists the item names and full statements for the feedback summarized in Table~\ref{tab:student_experience_results}.

\subsection{Feedback Summary}

Questionnaire items were summarized separately using the mean rating and the number and percentage of valid responses rated 4 or 5. All 39 participants provided valid responses for each item. We reported no composite score, CI, or significance test. These descriptive responses indicate students' perceptions of the interaction and its practice value, without establishing improvements in communication competence.

\begin{table*}[t]
\centering
\small
\setlength{\tabcolsep}{4pt}
\begin{tabular}{@{}cp{0.24\textwidth}p{0.68\textwidth}@{}}
\toprule
\textbf{Item}
& \textbf{Item name}
& \textbf{Questionnaire statement} \\
\midrule
1
& Contextual appropriateness
& The simulated patient provided reasonable and context-appropriate responses to my questions. \\

2
& Emotional realism
& The simulated patient's language and responses reflected realistic patient emotions and concerns. \\

3
& Communication insight
& The simulation helped me understand how communication strategies influence patient emotions, concerns, and responses. \\

4
& Practice value
& Interacting with the simulated patient helped me practise patient communication skills. \\

5
& Satisfaction and reuse
& Overall, I was satisfied with the learning experience provided by the platform and would be willing to use it again. \\
\bottomrule
\end{tabular}
\caption{Student questionnaire item names and full statements.}
\label{tab:student_questionnaire}
\end{table*}

\section{Baseline Comparison Results by Judge}
\label{sec:baseline_appendix}

All four LLM judges independently evaluated CPM-MultiAgent and the six baselines on 120 matched controlled cases. Tables~\ref{tab:baseline_gpt61sol}--\ref{tab:baseline_deepseekv4pro} report mean ratings and Holm-adjusted $p$ values for EUP, ARQ, BEQ, CR, and SQ. TC is inapplicable to single-trigger cases.

For each judge and criterion, two-sided paired $t$ tests compare CPM-MultiAgent with each baseline using within-case rating differences. Holm correction is applied across the six baseline comparisons separately for each judge and criterion. Table~\ref{tab:llm_judge_results} reports mean ALL, $\Delta_{\mathrm{ALL}}$, and paired-test $p$ values, with Holm correction across the six ALL contrasts separately within each judge. These composite comparisons are distinct from the criterion-level tests; ratings are not pooled across judges.

\begin{table*}[t]
\centering
\footnotesize
\setlength{\tabcolsep}{1.5pt}
\begin{tabular*}{\textwidth}{@{}>{\raggedright\arraybackslash}p{0.28\textwidth}@{\extracolsep{\fill}}*{12}{c}@{}}
\toprule
\multirow{2}{*}{\textbf{Method}}
& \multicolumn{3}{c}{\textbf{GPT-6.1 Sol}}
& \multicolumn{3}{c}{\textbf{Claude Opus 5.5}}
& \multicolumn{3}{c}{\textbf{DeepSeek-V4-Pro}}
& \multicolumn{3}{c}{\textbf{Gemini 3.1 Pro}} \\
\cmidrule(lr){2-4}\cmidrule(lr){5-7}
\cmidrule(lr){8-10}\cmidrule(lr){11-13}
& \textbf{ALL} & \textbf{$\Delta_{\mathrm{ALL}}$} & \textbf{Holm $p$}
& \textbf{ALL} & \textbf{$\Delta_{\mathrm{ALL}}$} & \textbf{Holm $p$}
& \textbf{ALL} & \textbf{$\Delta_{\mathrm{ALL}}$} & \textbf{Holm $p$}
& \textbf{ALL} & \textbf{$\Delta_{\mathrm{ALL}}$} & \textbf{Holm $p$} \\
\midrule
Zero-shot
& 4.30 & 0.51 & $<0.001$ & 3.43 & 0.55 & $<0.001$ & 4.45 & 0.26 & $<0.001$ & 3.02 & 0.73 & $<0.001$ \\
Few-shot
& 4.39 & 0.42 & $<0.001$ & 3.53 & 0.45 & $<0.001$ & 4.47 & 0.23 & $<0.001$ & 3.09 & 0.66 & $<0.001$ \\
Zero-shot CoT
& 4.54 & 0.27 & $<0.001$ & 3.74 & 0.24 & $<0.001$ & 4.51 & 0.20 & $<0.001$ & 3.42 & 0.33 & $<0.001$ \\
Few-shot CoT
& 4.64 & 0.18 & $<0.001$ & 3.87 & 0.11 & 0.006 & 4.61 & 0.10 & 0.102 & 3.52 & 0.23 & $<0.001$ \\
Self-consistency
& 4.74 & 0.07 & 0.051 & 3.90 & 0.08 & 0.039 & 4.62 & 0.09 & 0.146 & 3.55 & 0.20 & $<0.001$ \\
Self-Refine
& 4.58 & 0.24 & $<0.001$ & 3.90 & 0.08 & 0.021 & 4.63 & 0.07 & 0.146 & 3.55 & 0.20 & $<0.001$ \\
\midrule
CPM-MultiAgent
& 4.81 & 0.00 & -- & 3.98 & 0.00 & -- & 4.70 & 0.00 & -- & 3.75 & 0.00 & -- \\
\bottomrule
\end{tabular*}
\caption{Baseline comparisons on 120 matched controlled cases. Entries show mean ALL, paired differences, and Holm $p$. ALL is the equal-weight mean of EUP, ARQ, BEQ, CR, and SQ; $\Delta_{\mathrm{ALL}}$ is the mean paired CPM-MultiAgent-minus-baseline difference. Means and differences are computed before rounding. Holm $p$: two-sided paired $t$ tests adjusted across six contrasts per judge. A dash indicates no reference test.}
\label{tab:llm_judge_results}
\end{table*}

\begin{table*}[!t]

\centering
\footnotesize
\setlength{\tabcolsep}{2pt}
\renewcommand{\arraystretch}{1.10}
\begin{tabular*}{\textwidth}{@{}>{\raggedright\arraybackslash}p{.28\textwidth}@{\extracolsep{\fill}}*{10}{c}@{}}
\toprule
\multirow{2}{*}{\textbf{Method}} & \multicolumn{2}{c}{\textbf{EUP}} & \multicolumn{2}{c}{\textbf{ARQ}} & \multicolumn{2}{c}{\textbf{BEQ}} & \multicolumn{2}{c}{\textbf{CR}} & \multicolumn{2}{c}{\textbf{SQ}} \\
\cmidrule(lr){2-3}\cmidrule(lr){4-5}\cmidrule(lr){6-7}\cmidrule(lr){8-9}\cmidrule(lr){10-11}
 & \textbf{Mean} & \textbf{Holm $p$} & \textbf{Mean} & \textbf{Holm $p$} & \textbf{Mean} & \textbf{Holm $p$} & \textbf{Mean} & \textbf{Holm $p$} & \textbf{Mean} & \textbf{Holm $p$} \\
\midrule
Zero-shot & 4.475 & $<0.001$ & 4.183 & $<0.001$ & 4.192 & $<0.001$ & 4.408 & $<0.001$ & 4.233 & $<0.001$ \\
Few-shot & 4.767 & 0.014 & 4.483 & $<0.001$ & 3.950 & $<0.001$ & 4.467 & $<0.001$ & 4.300 & $<0.001$ \\
Zero-shot CoT & 4.683 & $<0.001$ & 4.517 & $<0.001$ & 4.358 & 0.012 & 4.675 & $<0.001$ & 4.467 & $<0.001$ \\
Few-shot CoT & 4.817 & 0.042 & 4.550 & $<0.001$ & 4.408 & 0.024 & 4.775 & 0.067 & 4.625 & 0.004 \\
Self-consistency & 4.850 & 0.090 & 4.850 & 0.117 & 4.508 & 0.391 & 4.800 & 0.088 & 4.708 & 0.096 \\
Self-Refine & 4.700 & $<0.001$ & 4.392 & $<0.001$ & 4.533 & 0.524 & 4.708 & 0.006 & 4.550 & $<0.001$ \\
\midrule
CPM-MultiAgent & 4.908 & -- & 4.917 & -- & 4.575 & -- & 4.867 & -- & 4.792 & -- \\
\bottomrule
\end{tabular*}
\caption{GPT-6.1 Sol baseline ratings on 120 matched controlled cases. Entries are means (1--5) and paired-test $p$ values, Holm-adjusted across six baselines per criterion. A dash marks the CPM-MultiAgent reference.}
\label{tab:baseline_gpt61sol}

\par\medskip

\centering
\footnotesize
\setlength{\tabcolsep}{2pt}
\renewcommand{\arraystretch}{1.10}
\begin{tabular*}{\textwidth}{@{}>{\raggedright\arraybackslash}p{.28\textwidth}@{\extracolsep{\fill}}*{10}{c}@{}}
\toprule
\multirow{2}{*}{\textbf{Method}} & \multicolumn{2}{c}{\textbf{EUP}} & \multicolumn{2}{c}{\textbf{ARQ}} & \multicolumn{2}{c}{\textbf{BEQ}} & \multicolumn{2}{c}{\textbf{CR}} & \multicolumn{2}{c}{\textbf{SQ}} \\
\cmidrule(lr){2-3}\cmidrule(lr){4-5}\cmidrule(lr){6-7}\cmidrule(lr){8-9}\cmidrule(lr){10-11}
 & \textbf{Mean} & \textbf{Holm $p$} & \textbf{Mean} & \textbf{Holm $p$} & \textbf{Mean} & \textbf{Holm $p$} & \textbf{Mean} & \textbf{Holm $p$} & \textbf{Mean} & \textbf{Holm $p$} \\
\midrule
Zero-shot & 3.733 & $<0.001$ & 3.333 & $<0.001$ & 3.183 & $<0.001$ & 3.525 & $<0.001$ & 3.350 & $<0.001$ \\
Few-shot & 3.858 & $<0.001$ & 3.375 & $<0.001$ & 3.333 & $<0.001$ & 3.625 & $<0.001$ & 3.475 & $<0.001$ \\
Zero-shot CoT & 3.942 & $<0.001$ & 3.708 & $<0.001$ & 3.533 & 0.003 & 3.825 & 0.002 & 3.692 & $<0.001$ \\
Few-shot CoT & 4.167 & 0.290 & 3.867 & 0.115 & 3.567 & 0.003 & 3.892 & 0.035 & 3.842 & 0.084 \\
Self-consistency & 4.167 & 0.290 & 3.992 & 0.614 & 3.625 & 0.058 & 3.883 & 0.035 & 3.842 & 0.084 \\
Self-Refine & 4.142 & 0.143 & 3.683 & $<0.001$ & 3.808 & 0.305 & 3.950 & 0.166 & 3.900 & 0.253 \\
\midrule
CPM-MultiAgent & 4.233 & -- & 3.967 & -- & 3.758 & -- & 3.992 & -- & 3.942 & -- \\
\bottomrule
\end{tabular*}
\caption{Claude Opus 5.5 baseline ratings on 120 matched controlled cases. Entries are means (1--5) and paired-test $p$ values, Holm-adjusted across six baselines per criterion. A dash marks the CPM-MultiAgent reference.}
\label{tab:baseline_opus55}

\end{table*}

\begin{table*}[!t]

\centering
\footnotesize
\setlength{\tabcolsep}{2pt}
\renewcommand{\arraystretch}{1.10}
\begin{tabular*}{\textwidth}{@{}>{\raggedright\arraybackslash}p{.28\textwidth}@{\extracolsep{\fill}}*{10}{c}@{}}
\toprule
\multirow{2}{*}{\textbf{Method}} & \multicolumn{2}{c}{\textbf{EUP}} & \multicolumn{2}{c}{\textbf{ARQ}} & \multicolumn{2}{c}{\textbf{BEQ}} & \multicolumn{2}{c}{\textbf{CR}} & \multicolumn{2}{c}{\textbf{SQ}} \\
\cmidrule(lr){2-3}\cmidrule(lr){4-5}\cmidrule(lr){6-7}\cmidrule(lr){8-9}\cmidrule(lr){10-11}
 & \textbf{Mean} & \textbf{Holm $p$} & \textbf{Mean} & \textbf{Holm $p$} & \textbf{Mean} & \textbf{Holm $p$} & \textbf{Mean} & \textbf{Holm $p$} & \textbf{Mean} & \textbf{Holm $p$} \\
\midrule
Zero-shot & 3.433 & $<0.001$ & 2.808 & $<0.001$ & 2.767 & $<0.001$ & 3.175 & $<0.001$ & 2.925 & $<0.001$ \\
Few-shot & 3.483 & $<0.001$ & 2.767 & $<0.001$ & 2.933 & $<0.001$ & 3.250 & $<0.001$ & 2.992 & $<0.001$ \\
Zero-shot CoT & 3.825 & 0.633 & 3.058 & $<0.001$ & 3.225 & $<0.001$ & 3.667 & $<0.001$ & 3.325 & $<0.001$ \\
Few-shot CoT & 3.900 & 1.000 & 3.317 & $<0.001$ & 3.225 & $<0.001$ & 3.700 & $<0.001$ & 3.433 & $<0.001$ \\
Self-consistency & 3.900 & 1.000 & 3.325 & $<0.001$ & 3.317 & $<0.001$ & 3.683 & $<0.001$ & 3.508 & $<0.001$ \\
Self-Refine & 3.842 & 0.872 & 3.258 & $<0.001$ & 3.425 & 0.010 & 3.717 & 0.003 & 3.492 & $<0.001$ \\
\midrule
CPM-MultiAgent & 3.892 & -- & 3.633 & -- & 3.592 & -- & 3.875 & -- & 3.742 & -- \\
\bottomrule
\end{tabular*}
\caption{Gemini 3.1 Pro baseline ratings on 120 matched controlled cases. Entries are means (1--5) and paired-test $p$ values, Holm-adjusted across six baselines per criterion. A dash marks the CPM-MultiAgent reference.}
\label{tab:baseline_gemini31pro}

\par\medskip

\centering
\footnotesize
\setlength{\tabcolsep}{2pt}
\renewcommand{\arraystretch}{1.10}
\begin{tabular*}{\textwidth}{@{}>{\raggedright\arraybackslash}p{.28\textwidth}@{\extracolsep{\fill}}*{10}{c}@{}}
\toprule
\multirow{2}{*}{\textbf{Method}} & \multicolumn{2}{c}{\textbf{EUP}} & \multicolumn{2}{c}{\textbf{ARQ}} & \multicolumn{2}{c}{\textbf{BEQ}} & \multicolumn{2}{c}{\textbf{CR}} & \multicolumn{2}{c}{\textbf{SQ}} \\
\cmidrule(lr){2-3}\cmidrule(lr){4-5}\cmidrule(lr){6-7}\cmidrule(lr){8-9}\cmidrule(lr){10-11}
 & \textbf{Mean} & \textbf{Holm $p$} & \textbf{Mean} & \textbf{Holm $p$} & \textbf{Mean} & \textbf{Holm $p$} & \textbf{Mean} & \textbf{Holm $p$} & \textbf{Mean} & \textbf{Holm $p$} \\
\midrule
Zero-shot & 4.558 & $<0.001$ & 4.317 & 0.002 & 4.383 & 0.003 & 4.550 & 0.008 & 4.417 & $<0.001$ \\
Few-shot & 4.650 & 0.011 & 4.400 & 0.036 & 4.342 & $<0.001$ & 4.558 & 0.002 & 4.417 & $<0.001$ \\
Zero-shot CoT & 4.592 & $<0.001$ & 4.350 & 0.006 & 4.550 & 0.416 & 4.592 & 0.037 & 4.442 & $<0.001$ \\
Few-shot CoT & 4.708 & 0.068 & 4.508 & 0.616 & 4.592 & 0.588 & 4.675 & 0.247 & 4.550 & 0.105 \\
Self-consistency & 4.692 & 0.068 & 4.600 & 1.000 & 4.558 & 0.493 & 4.667 & 0.247 & 4.567 & 0.107 \\
Self-Refine & 4.683 & 0.015 & 4.600 & 1.000 & 4.675 & 0.899 & 4.650 & 0.206 & 4.550 & 0.105 \\
\midrule
CPM-MultiAgent & 4.808 & -- & 4.600 & -- & 4.667 & -- & 4.758 & -- & 4.675 & -- \\
\bottomrule
\end{tabular*}
\caption{DeepSeek-V4-Pro baseline ratings on 120 matched controlled cases. Entries are means (1--5) and paired-test $p$ values, Holm-adjusted across six baselines per criterion. A dash marks the CPM-MultiAgent reference.}
\label{tab:baseline_deepseekv4pro}

\end{table*}

\section{Detailed Analysis of Ablation Study}
\label{sec:ablation_appendix}

Each judge evaluated the full CPM-MultiAgent system and the seven component-removal variants on the same 120 controlled single-trigger cases. Tables~\ref{tab:ablation_gpt61sol}--\ref{tab:ablation_deepseekv4pro} report mean ratings and Holm-adjusted $p$ values for EUP, ARQ, BEQ, CR, and SQ separately for each judge. TC is inapplicable to this task.

Two-sided paired $t$ tests compare the full system with each variant using within-case rating differences. Holm correction is applied across the seven removals separately for each judge and criterion, as a family distinct from the baseline comparisons. Table~\ref{tab:ablation_results} reports mean ALL, $\Delta_{\mathrm{ALL}}$, and paired-test $p$ values, with Holm correction across the seven ALL contrasts separately within each judge.

\begin{table*}[t]
\centering
\footnotesize
\setlength{\tabcolsep}{1.5pt}
\begin{tabular*}{\textwidth}{@{}>{\raggedright\arraybackslash}p{0.28\textwidth}@{\extracolsep{\fill}}*{12}{c}@{}}
\toprule
\multirow{2}{*}{\textbf{System variant}}
& \multicolumn{3}{c}{\textbf{GPT-6.1 Sol}}
& \multicolumn{3}{c}{\textbf{Claude Opus 5.5}}
& \multicolumn{3}{c}{\textbf{DeepSeek-V4-Pro}}
& \multicolumn{3}{c}{\textbf{Gemini 3.1 Pro}} \\
\cmidrule(lr){2-4}\cmidrule(lr){5-7}
\cmidrule(lr){8-10}\cmidrule(lr){11-13}
& \textbf{ALL} & \textbf{$\Delta_{\mathrm{ALL}}$} & \textbf{Holm $p$}
& \textbf{ALL} & \textbf{$\Delta_{\mathrm{ALL}}$} & \textbf{Holm $p$}
& \textbf{ALL} & \textbf{$\Delta_{\mathrm{ALL}}$} & \textbf{Holm $p$}
& \textbf{ALL} & \textbf{$\Delta_{\mathrm{ALL}}$} & \textbf{Holm $p$} \\
\midrule
Full CPM-MultiAgent
& 4.81 & 0.00 & -- & 3.98 & 0.00 & -- & 4.70 & 0.00 & -- & 3.75 & 0.00 & -- \\
w/o Trigger Analysis
& 4.59 & 0.23 & $<0.001$ & 3.79 & 0.19 & $<0.001$ & 4.56 & 0.14 & 0.026 & 3.38 & 0.37 & $<0.001$ \\
w/o Relevance Check
& 4.58 & 0.24 & $<0.001$ & 3.88 & 0.10 & 0.021 & 4.58 & 0.13 & 0.035 & 3.45 & 0.29 & $<0.001$ \\
w/o Implication Check
& 4.54 & 0.28 & $<0.001$ & 3.87 & 0.11 & $<0.001$ & 4.49 & 0.21 & $<0.001$ & 3.38 & 0.37 & $<0.001$ \\
w/o Coping Potential Check
& 4.63 & 0.19 & $<0.001$ & 3.87 & 0.11 & 0.002 & 4.59 & 0.11 & 0.035 & 3.40 & 0.34 & $<0.001$ \\
w/o Normative Significance Check
& 4.73 & 0.09 & 0.017 & 3.91 & 0.07 & 0.039 & 4.63 & 0.07 & 0.132 & 3.51 & 0.24 & $<0.001$ \\
w/o Cross-Appraisal Peer Review
& 4.65 & 0.16 & $<0.001$ & 3.90 & 0.08 & 0.021 & 4.56 & 0.14 & 0.026 & 3.48 & 0.27 & $<0.001$ \\
w/o Consistency Critique
& 4.69 & 0.13 & 0.010 & 3.89 & 0.09 & 0.021 & 4.62 & 0.08 & 0.132 & 3.42 & 0.33 & $<0.001$ \\
\bottomrule
\end{tabular*}
\caption{Ablation comparisons on 120 matched controlled cases. Entries show mean ALL, paired differences, and Holm $p$. ALL is the equal-weight mean of EUP, ARQ, BEQ, CR, and SQ; $\Delta_{\mathrm{ALL}}$ is the mean paired full-system-minus-variant difference. Means and differences are computed before rounding. Holm $p$: two-sided paired $t$ tests adjusted across seven contrasts per judge. A dash indicates no reference test.}
\label{tab:ablation_results}
\end{table*}

\begin{table*}[!t]

\centering
\footnotesize
\setlength{\tabcolsep}{2pt}
\renewcommand{\arraystretch}{1.10}
\begin{tabular*}{\textwidth}{@{}>{\raggedright\arraybackslash}p{.28\textwidth}@{\extracolsep{\fill}}*{10}{c}@{}}
\toprule
\multirow{2}{*}{\textbf{System variant}} & \multicolumn{2}{c}{\textbf{EUP}} & \multicolumn{2}{c}{\textbf{ARQ}} & \multicolumn{2}{c}{\textbf{BEQ}} & \multicolumn{2}{c}{\textbf{CR}} & \multicolumn{2}{c}{\textbf{SQ}} \\
\cmidrule(lr){2-3}\cmidrule(lr){4-5}\cmidrule(lr){6-7}\cmidrule(lr){8-9}\cmidrule(lr){10-11}
 & \textbf{Mean} & \textbf{Holm $p$} & \textbf{Mean} & \textbf{Holm $p$} & \textbf{Mean} & \textbf{Holm $p$} & \textbf{Mean} & \textbf{Holm $p$} & \textbf{Mean} & \textbf{Holm $p$} \\
\midrule
Full CPM-MultiAgent & 4.908 & -- & 4.917 & -- & 4.575 & -- & 4.867 & -- & 4.792 & -- \\
\midrule
w/o Trigger Analysis & 4.742 & $<0.001$ & 4.367 & $<0.001$ & 4.508 & 0.346 & 4.725 & 0.017 & 4.592 & 0.001 \\
w/o Relevance Check & 4.783 & 0.043 & 4.508 & $<0.001$ & 4.367 & 0.016 & 4.708 & 0.017 & 4.517 & $<0.001$ \\
w/o Implication Check & 4.742 & 0.001 & 4.267 & $<0.001$ & 4.483 & 0.346 & 4.717 & 0.017 & 4.475 & $<0.001$ \\
w/o Coping Potential Check & 4.800 & 0.031 & 4.567 & $<0.001$ & 4.442 & 0.192 & 4.750 & 0.051 & 4.575 & 0.001 \\
w/o Normative Significance Check & 4.867 & 0.594 & 4.842 & 0.083 & 4.475 & 0.307 & 4.758 & 0.082 & 4.692 & 0.017 \\
w/o Cross-Appraisal Peer Review & 4.867 & 0.594 & 4.517 & $<0.001$ & 4.475 & 0.346 & 4.817 & 0.367 & 4.567 & $<0.001$ \\
w/o Consistency Critique & 4.858 & 0.594 & 4.708 & $<0.001$ & 4.425 & 0.157 & 4.800 & 0.367 & 4.633 & 0.017 \\
\bottomrule
\end{tabular*}
\caption{GPT-6.1 Sol ablation ratings on 120 matched controlled cases. Entries are means (1--5) and paired-test $p$ values, Holm-adjusted across seven removals per criterion. A dash marks the full-system reference.}
\label{tab:ablation_gpt61sol}

\par\medskip

\centering
\footnotesize
\setlength{\tabcolsep}{2pt}
\renewcommand{\arraystretch}{1.10}
\begin{tabular*}{\textwidth}{@{}>{\raggedright\arraybackslash}p{.28\textwidth}@{\extracolsep{\fill}}*{10}{c}@{}}
\toprule
\multirow{2}{*}{\textbf{System variant}} & \multicolumn{2}{c}{\textbf{EUP}} & \multicolumn{2}{c}{\textbf{ARQ}} & \multicolumn{2}{c}{\textbf{BEQ}} & \multicolumn{2}{c}{\textbf{CR}} & \multicolumn{2}{c}{\textbf{SQ}} \\
\cmidrule(lr){2-3}\cmidrule(lr){4-5}\cmidrule(lr){6-7}\cmidrule(lr){8-9}\cmidrule(lr){10-11}
 & \textbf{Mean} & \textbf{Holm $p$} & \textbf{Mean} & \textbf{Holm $p$} & \textbf{Mean} & \textbf{Holm $p$} & \textbf{Mean} & \textbf{Holm $p$} & \textbf{Mean} & \textbf{Holm $p$} \\
\midrule
Full CPM-MultiAgent & 4.233 & -- & 3.967 & -- & 3.758 & -- & 3.992 & -- & 3.942 & -- \\
\midrule
w/o Trigger Analysis & 4.025 & $<0.001$ & 3.533 & $<0.001$ & 3.758 & 1.000 & 3.842 & 0.006 & 3.800 & 0.042 \\
w/o Relevance Check & 4.083 & 0.014 & 3.775 & 0.004 & 3.742 & 1.000 & 3.950 & 0.333 & 3.867 & 0.215 \\
w/o Implication Check & 4.125 & 0.031 & 3.642 & $<0.001$ & 3.767 & 1.000 & 3.942 & 0.333 & 3.850 & 0.043 \\
w/o Coping Potential Check & 4.142 & 0.109 & 3.817 & 0.020 & 3.633 & 0.233 & 3.892 & 0.060 & 3.842 & 0.090 \\
w/o Normative Significance Check & 4.192 & 0.338 & 3.908 & 0.552 & 3.633 & 0.233 & 3.933 & 0.333 & 3.883 & 0.215 \\
w/o Cross-Appraisal Peer Review & 4.142 & 0.122 & 3.858 & 0.141 & 3.708 & 1.000 & 3.917 & 0.246 & 3.850 & 0.109 \\
w/o Consistency Critique & 4.150 & 0.122 & 3.908 & 0.552 & 3.558 & 0.006 & 3.933 & 0.333 & 3.875 & 0.215 \\
\bottomrule
\end{tabular*}
\caption{Claude Opus 5.5 ablation ratings on 120 matched controlled cases. Entries are means (1--5) and paired-test $p$ values, Holm-adjusted across seven removals per criterion. A dash marks the full-system reference.}
\label{tab:ablation_opus55}

\end{table*}

\begin{table*}[!t]

\centering
\footnotesize
\setlength{\tabcolsep}{2pt}
\renewcommand{\arraystretch}{1.10}
\begin{tabular*}{\textwidth}{@{}>{\raggedright\arraybackslash}p{.28\textwidth}@{\extracolsep{\fill}}*{10}{c}@{}}
\toprule
\multirow{2}{*}{\textbf{System variant}} & \multicolumn{2}{c}{\textbf{EUP}} & \multicolumn{2}{c}{\textbf{ARQ}} & \multicolumn{2}{c}{\textbf{BEQ}} & \multicolumn{2}{c}{\textbf{CR}} & \multicolumn{2}{c}{\textbf{SQ}} \\
\cmidrule(lr){2-3}\cmidrule(lr){4-5}\cmidrule(lr){6-7}\cmidrule(lr){8-9}\cmidrule(lr){10-11}
 & \textbf{Mean} & \textbf{Holm $p$} & \textbf{Mean} & \textbf{Holm $p$} & \textbf{Mean} & \textbf{Holm $p$} & \textbf{Mean} & \textbf{Holm $p$} & \textbf{Mean} & \textbf{Holm $p$} \\
\midrule
Full CPM-MultiAgent & 3.892 & -- & 3.633 & -- & 3.592 & -- & 3.875 & -- & 3.742 & -- \\
\midrule
w/o Trigger Analysis & 3.733 & 0.025 & 3.100 & $<0.001$ & 3.183 & $<0.001$ & 3.608 & $<0.001$ & 3.267 & $<0.001$ \\
w/o Relevance Check & 3.817 & 0.643 & 3.275 & $<0.001$ & 3.183 & $<0.001$ & 3.608 & $<0.001$ & 3.383 & $<0.001$ \\
w/o Implication Check & 3.733 & 0.011 & 2.792 & $<0.001$ & 3.383 & 0.002 & 3.625 & $<0.001$ & 3.358 & $<0.001$ \\
w/o Coping Potential Check & 3.775 & 0.171 & 3.133 & $<0.001$ & 3.125 & $<0.001$ & 3.667 & $<0.001$ & 3.317 & $<0.001$ \\
w/o Normative Significance Check & 3.900 & 1.000 & 3.283 & $<0.001$ & 3.233 & $<0.001$ & 3.650 & $<0.001$ & 3.483 & $<0.001$ \\
w/o Cross-Appraisal Peer Review & 3.883 & 1.000 & 3.150 & $<0.001$ & 3.275 & $<0.001$ & 3.675 & $<0.001$ & 3.392 & $<0.001$ \\
w/o Consistency Critique & 3.867 & 1.000 & 3.133 & $<0.001$ & 3.183 & $<0.001$ & 3.617 & $<0.001$ & 3.308 & $<0.001$ \\
\bottomrule
\end{tabular*}
\caption{Gemini 3.1 Pro ablation ratings on 120 matched controlled cases. Entries are means (1--5) and paired-test $p$ values, Holm-adjusted across seven removals per criterion. A dash marks the full-system reference.}
\label{tab:ablation_gemini31pro}

\par\medskip

\centering
\footnotesize
\setlength{\tabcolsep}{2pt}
\renewcommand{\arraystretch}{1.10}
\begin{tabular*}{\textwidth}{@{}>{\raggedright\arraybackslash}p{.28\textwidth}@{\extracolsep{\fill}}*{10}{c}@{}}
\toprule
\multirow{2}{*}{\textbf{System variant}} & \multicolumn{2}{c}{\textbf{EUP}} & \multicolumn{2}{c}{\textbf{ARQ}} & \multicolumn{2}{c}{\textbf{BEQ}} & \multicolumn{2}{c}{\textbf{CR}} & \multicolumn{2}{c}{\textbf{SQ}} \\
\cmidrule(lr){2-3}\cmidrule(lr){4-5}\cmidrule(lr){6-7}\cmidrule(lr){8-9}\cmidrule(lr){10-11}
 & \textbf{Mean} & \textbf{Holm $p$} & \textbf{Mean} & \textbf{Holm $p$} & \textbf{Mean} & \textbf{Holm $p$} & \textbf{Mean} & \textbf{Holm $p$} & \textbf{Mean} & \textbf{Holm $p$} \\
\midrule
Full CPM-MultiAgent & 4.808 & -- & 4.600 & -- & 4.667 & -- & 4.758 & -- & 4.675 & -- \\
\midrule
w/o Trigger Analysis & 4.575 & $<0.001$ & 4.458 & 0.341 & 4.625 & 1.000 & 4.658 & 0.226 & 4.500 & 0.043 \\
w/o Relevance Check & 4.583 & 0.001 & 4.542 & 0.921 & 4.642 & 1.000 & 4.625 & 0.122 & 4.492 & 0.018 \\
w/o Implication Check & 4.583 & $<0.001$ & 4.217 & $<0.001$ & 4.650 & 1.000 & 4.583 & 0.013 & 4.433 & $<0.001$ \\
w/o Coping Potential Check & 4.575 & $<0.001$ & 4.483 & 0.397 & 4.683 & 1.000 & 4.692 & 0.226 & 4.525 & 0.043 \\
w/o Normative Significance Check & 4.733 & 0.213 & 4.517 & 0.680 & 4.625 & 1.000 & 4.658 & 0.226 & 4.608 & 0.218 \\
w/o Cross-Appraisal Peer Review & 4.700 & 0.069 & 4.425 & 0.123 & 4.558 & 1.000 & 4.625 & 0.122 & 4.508 & 0.043 \\
w/o Consistency Critique & 4.742 & 0.229 & 4.600 & 1.000 & 4.550 & 0.922 & 4.642 & 0.226 & 4.558 & 0.132 \\
\bottomrule
\end{tabular*}
\caption{DeepSeek-V4-Pro ablation ratings on 120 matched controlled cases. Entries are means (1--5) and paired-test $p$ values, Holm-adjusted across seven removals per criterion. A dash marks the full-system reference.}
\label{tab:ablation_deepseekv4pro}

\end{table*}

\section{Backbone Robustness Results by Judge}
\label{sec:backbone_appendix}

The generation backbone was varied while the CPM-MultiAgent architecture, evaluation inputs, and four judge models remained fixed. Each judge evaluated outputs from all six backbones on the same 120 controlled single-trigger cases. EUP, ARQ, BEQ, CR, and SQ were assessed; TC was inapplicable to this task.

Ratings were summarized separately for each judge using the case-level bootstrap procedure in Appendix~\ref{sec:llm_eval_procedure}. The backbone comparisons are descriptive.

\begin{table*}[!t]

\centering
\footnotesize
\setlength{\tabcolsep}{3pt}
\renewcommand{\arraystretch}{1.10}
\begin{tabular*}{\textwidth}{@{}>{\raggedright\arraybackslash}p{0.28\textwidth}@{\extracolsep{\fill}}*{5}{c}@{}}
\toprule
\textbf{Generation backbone} & \textbf{EUP} & \textbf{ARQ} & \textbf{BEQ} & \textbf{CR} & \textbf{SQ} \\
\midrule
GPT-5.6 Luna & $4.908 \pm 0.317$ & $4.917 \pm 0.278$ & $4.575 \pm 0.617$ & $4.867 \pm 0.365$ & $4.792 \pm 0.428$ \\
95\% CI & $[4.850,\,4.958]$ & $[4.867,\,4.958]$ & $[4.458,\,4.683]$ & $[4.800,\,4.925]$ & $[4.717,\,4.867]$ \\
\addlinespace[2pt]
DeepSeek-V4.1-Flash & $4.908 \pm 0.317$ & $4.908 \pm 0.343$ & $4.508 \pm 0.674$ & $4.825 \pm 0.479$ & $4.800 \pm 0.442$ \\
95\% CI & $[4.850,\,4.958]$ & $[4.842,\,4.967]$ & $[4.383,\,4.625]$ & $[4.733,\,4.908]$ & $[4.717,\,4.875]$ \\
\addlinespace[2pt]
Claude Haiku 4.5 & $4.883 \pm 0.393$ & $4.883 \pm 0.371$ & $4.483 \pm 0.698$ & $4.808 \pm 0.473$ & $4.708 \pm 0.556$ \\
95\% CI & $[4.808,\,4.950]$ & $[4.817,\,4.942]$ & $[4.358,\,4.600]$ & $[4.717,\,4.892]$ & $[4.608,\,4.800]$ \\
\addlinespace[2pt]
Gemini 3.5 Flash-Lite & $4.892 \pm 0.312$ & $4.858 \pm 0.373$ & $4.467 \pm 0.593$ & $4.800 \pm 0.460$ & $4.742 \pm 0.458$ \\
95\% CI & $[4.833,\,4.942]$ & $[4.792,\,4.917]$ & $[4.358,\,4.575]$ & $[4.717,\,4.875]$ & $[4.658,\,4.825]$ \\
\addlinespace[2pt]
GPT-OSS 120B & $4.358 \pm 0.708$ & $4.117 \pm 0.758$ & $4.017 \pm 0.778$ & $4.325 \pm 0.700$ & $4.108 \pm 0.719$ \\
95\% CI & $[4.233,\,4.483]$ & $[3.975,\,4.250]$ & $[3.875,\,4.158]$ & $[4.200,\,4.450]$ & $[3.983,\,4.233]$ \\
\addlinespace[2pt]
Gemma 4 26B A4B & $4.058 \pm 0.823$ & $3.783 \pm 0.881$ & $4.142 \pm 0.598$ & $4.100 \pm 0.679$ & $3.950 \pm 0.696$ \\
95\% CI & $[3.917,\,4.200]$ & $[3.633,\,3.942]$ & $[4.033,\,4.250]$ & $[3.983,\,4.225]$ & $[3.833,\,4.075]$ \\
\bottomrule
\end{tabular*}
\caption{Backbone ratings by GPT-6.1 Sol on 120 matched controlled cases. Upper rows: mean $\pm$ sample SD. Lower rows: 95\% percentile-bootstrap CIs (10,000 case-level resamples).}
\label{tab:backbone_gpt61sol}

\par\medskip

\centering
\footnotesize
\setlength{\tabcolsep}{3pt}
\renewcommand{\arraystretch}{1.10}
\begin{tabular*}{\textwidth}{@{}>{\raggedright\arraybackslash}p{0.28\textwidth}@{\extracolsep{\fill}}*{5}{c}@{}}
\toprule
\textbf{Generation backbone} & \textbf{EUP} & \textbf{ARQ} & \textbf{BEQ} & \textbf{CR} & \textbf{SQ} \\
\midrule
GPT-5.6 Luna & $4.233 \pm 0.480$ & $3.967 \pm 0.484$ & $3.758 \pm 0.467$ & $3.992 \pm 0.330$ & $3.942 \pm 0.325$ \\
95\% CI & $[4.150,\,4.317]$ & $[3.883,\,4.050]$ & $[3.675,\,3.842]$ & $[3.933,\,4.050]$ & $[3.883,\,4.000]$ \\
\addlinespace[2pt]
DeepSeek-V4.1-Flash & $4.183 \pm 0.518$ & $3.842 \pm 0.518$ & $3.692 \pm 0.499$ & $3.958 \pm 0.353$ & $3.892 \pm 0.406$ \\
95\% CI & $[4.092,\,4.275]$ & $[3.750,\,3.933]$ & $[3.600,\,3.775]$ & $[3.892,\,4.017]$ & $[3.817,\,3.958]$ \\
\addlinespace[2pt]
Claude Haiku 4.5 & $4.183 \pm 0.430$ & $3.908 \pm 0.367$ & $3.792 \pm 0.466$ & $3.950 \pm 0.339$ & $3.900 \pm 0.353$ \\
95\% CI & $[4.108,\,4.258]$ & $[3.842,\,3.975]$ & $[3.708,\,3.875]$ & $[3.883,\,4.008]$ & $[3.833,\,3.958]$ \\
\addlinespace[2pt]
Gemini 3.5 Flash-Lite & $4.183 \pm 0.485$ & $3.967 \pm 0.448$ & $3.625 \pm 0.536$ & $3.917 \pm 0.357$ & $3.883 \pm 0.371$ \\
95\% CI & $[4.100,\,4.275]$ & $[3.883,\,4.042]$ & $[3.525,\,3.717]$ & $[3.850,\,3.975]$ & $[3.817,\,3.950]$ \\
\addlinespace[2pt]
GPT-OSS 120B & $3.808 \pm 0.569$ & $3.325 \pm 0.624$ & $2.908 \pm 0.686$ & $3.408 \pm 0.601$ & $3.275 \pm 0.518$ \\
95\% CI & $[3.708,\,3.908]$ & $[3.217,\,3.433]$ & $[2.783,\,3.033]$ & $[3.300,\,3.517]$ & $[3.183,\,3.367]$ \\
\addlinespace[2pt]
Gemma 4 26B A4B & $3.433 \pm 0.877$ & $2.992 \pm 0.704$ & $3.483 \pm 0.622$ & $3.417 \pm 0.681$ & $3.250 \pm 0.701$ \\
95\% CI & $[3.283,\,3.592]$ & $[2.867,\,3.117]$ & $[3.375,\,3.592]$ & $[3.300,\,3.533]$ & $[3.125,\,3.375]$ \\
\bottomrule
\end{tabular*}
\caption{Backbone ratings by Claude Opus 5.5 on 120 matched controlled cases. Upper rows: mean $\pm$ sample SD. Lower rows: 95\% percentile-bootstrap CIs (10,000 case-level resamples).}
\label{tab:backbone_opus55}

\end{table*}

\begin{table*}[!t]

\centering
\footnotesize
\setlength{\tabcolsep}{3pt}
\renewcommand{\arraystretch}{1.10}
\begin{tabular*}{\textwidth}{@{}>{\raggedright\arraybackslash}p{0.28\textwidth}@{\extracolsep{\fill}}*{5}{c}@{}}
\toprule
\textbf{Generation backbone} & \textbf{EUP} & \textbf{ARQ} & \textbf{BEQ} & \textbf{CR} & \textbf{SQ} \\
\midrule
GPT-5.6 Luna & $3.892 \pm 0.384$ & $3.633 \pm 0.517$ & $3.592 \pm 0.527$ & $3.875 \pm 0.357$ & $3.742 \pm 0.440$ \\
95\% CI & $[3.825,\,3.958]$ & $[3.542,\,3.725]$ & $[3.500,\,3.683]$ & $[3.808,\,3.933]$ & $[3.658,\,3.817]$ \\
\addlinespace[2pt]
DeepSeek-V4.1-Flash & $3.900 \pm 0.376$ & $3.308 \pm 0.531$ & $3.508 \pm 0.565$ & $3.808 \pm 0.436$ & $3.567 \pm 0.530$ \\
95\% CI & $[3.825,\,3.967]$ & $[3.217,\,3.400]$ & $[3.408,\,3.608]$ & $[3.725,\,3.883]$ & $[3.467,\,3.658]$ \\
\addlinespace[2pt]
Claude Haiku 4.5 & $3.850 \pm 0.423$ & $3.283 \pm 0.471$ & $3.458 \pm 0.634$ & $3.775 \pm 0.458$ & $3.508 \pm 0.580$ \\
95\% CI & $[3.775,\,3.925]$ & $[3.200,\,3.367]$ & $[3.342,\,3.567]$ & $[3.692,\,3.850]$ & $[3.408,\,3.608]$ \\
\addlinespace[2pt]
Gemini 3.5 Flash-Lite & $3.783 \pm 0.488$ & $3.192 \pm 0.490$ & $3.517 \pm 0.534$ & $3.808 \pm 0.416$ & $3.483 \pm 0.550$ \\
95\% CI & $[3.692,\,3.867]$ & $[3.100,\,3.275]$ & $[3.417,\,3.608]$ & $[3.733,\,3.875]$ & $[3.383,\,3.583]$ \\
\addlinespace[2pt]
GPT-OSS 120B & $3.225 \pm 0.761$ & $2.567 \pm 0.546$ & $3.050 \pm 0.592$ & $3.167 \pm 0.626$ & $2.925 \pm 0.568$ \\
95\% CI & $[3.092,\,3.358]$ & $[2.467,\,2.667]$ & $[2.950,\,3.158]$ & $[3.058,\,3.275]$ & $[2.825,\,3.025]$ \\
\addlinespace[2pt]
Gemma 4 26B A4B & $3.250 \pm 0.583$ & $2.375 \pm 0.486$ & $2.442 \pm 0.547$ & $2.867 \pm 0.466$ & $2.542 \pm 0.517$ \\
95\% CI & $[3.142,\,3.358]$ & $[2.292,\,2.467]$ & $[2.342,\,2.542]$ & $[2.783,\,2.950]$ & $[2.450,\,2.633]$ \\
\bottomrule
\end{tabular*}
\caption{Backbone ratings by Gemini 3.1 Pro on 120 matched controlled cases. Upper rows: mean $\pm$ sample SD. Lower rows: 95\% percentile-bootstrap CIs (10,000 case-level resamples).}
\label{tab:backbone_gemini31pro}

\par\medskip

\centering
\footnotesize
\setlength{\tabcolsep}{3pt}
\renewcommand{\arraystretch}{1.10}
\begin{tabular*}{\textwidth}{@{}>{\raggedright\arraybackslash}p{0.28\textwidth}@{\extracolsep{\fill}}*{5}{c}@{}}
\toprule
\textbf{Generation backbone} & \textbf{EUP} & \textbf{ARQ} & \textbf{BEQ} & \textbf{CR} & \textbf{SQ} \\
\midrule
GPT-5.6 Luna & $4.808 \pm 0.416$ & $4.600 \pm 0.525$ & $4.667 \pm 0.613$ & $4.758 \pm 0.502$ & $4.675 \pm 0.521$ \\
95\% CI & $[4.733,\,4.883]$ & $[4.508,\,4.692]$ & $[4.550,\,4.775]$ & $[4.667,\,4.842]$ & $[4.575,\,4.767]$ \\
\addlinespace[2pt]
DeepSeek-V4.1-Flash & $4.700 \pm 0.544$ & $4.575 \pm 0.575$ & $4.708 \pm 0.509$ & $4.750 \pm 0.472$ & $4.617 \pm 0.553$ \\
95\% CI & $[4.600,\,4.792]$ & $[4.475,\,4.675]$ & $[4.617,\,4.792]$ & $[4.658,\,4.833]$ & $[4.517,\,4.717]$ \\
\addlinespace[2pt]
Claude Haiku 4.5 & $4.667 \pm 0.473$ & $4.525 \pm 0.608$ & $4.700 \pm 0.512$ & $4.758 \pm 0.485$ & $4.583 \pm 0.512$ \\
95\% CI & $[4.583,\,4.750]$ & $[4.417,\,4.633]$ & $[4.608,\,4.792]$ & $[4.667,\,4.842]$ & $[4.492,\,4.675]$ \\
\addlinespace[2pt]
Gemini 3.5 Flash-Lite & $4.700 \pm 0.495$ & $4.467 \pm 0.634$ & $4.667 \pm 0.570$ & $4.708 \pm 0.585$ & $4.625 \pm 0.551$ \\
95\% CI & $[4.608,\,4.783]$ & $[4.350,\,4.575]$ & $[4.558,\,4.758]$ & $[4.600,\,4.808]$ & $[4.525,\,4.717]$ \\
\addlinespace[2pt]
GPT-OSS 120B & $4.325 \pm 0.724$ & $4.342 \pm 0.716$ & $4.592 \pm 0.628$ & $4.525 \pm 0.635$ & $4.350 \pm 0.617$ \\
95\% CI & $[4.192,\,4.450]$ & $[4.217,\,4.467]$ & $[4.475,\,4.700]$ & $[4.408,\,4.633]$ & $[4.242,\,4.458]$ \\
\addlinespace[2pt]
Gemma 4 26B A4B & $4.483 \pm 0.594$ & $3.908 \pm 0.580$ & $4.425 \pm 0.774$ & $4.567 \pm 0.546$ & $4.233 \pm 0.576$ \\
95\% CI & $[4.375,\,4.592]$ & $[3.808,\,4.008]$ & $[4.283,\,4.558]$ & $[4.467,\,4.667]$ & $[4.133,\,4.333]$ \\
\bottomrule
\end{tabular*}
\caption{Backbone ratings by DeepSeek-V4-Pro on 120 matched controlled cases. Upper rows: mean $\pm$ sample SD. Lower rows: 95\% percentile-bootstrap CIs (10,000 case-level resamples).}
\label{tab:backbone_deepseekv4pro}

\end{table*}

\end{document}